\documentstyle[12pt,psfig]{article}
\newcommand{\mysection}[1]{\setcounter{equation}{0}\section{#1}}

\newcommand{\nc}{\newcommand}
\nc{\beq}{\begin{equation}} \nc{\eeq}{\end{equation}}
\nc{\beqa}{\begin{eqnarray}} \nc{\eeqa}{\end{eqnarray}}
\nc{\lsim}{\begin{array}{c}\,\sim\vspace{-21pt}\\< \end{array}}
\nc{\gsim}{\begin{array}{c}\sim\vspace{-21pt}\\> \end{array}}

\nc{\scR}{{\cal R}}
\nc{\scL}{{\cal L}}
\nc{\al}{\alpha}
\nc{\ald}{\dot{\alpha}}
\nc{\lam}{\lambda}
\nc{\nud}{\dot{\nu}}
\nc{\lamd}{\dot{\lam}}
\nc{\bed}{\dot{\beta}}

\nc{\Q}[2]{Q_{#1#2}}
\nc{\R}[2]{R_{#1#2}}
\nc{\Y}[2]{Y_{#1#2}}
\nc{\V}[2]{V_{#1#2}}
\nc{\q}[2]{q_{#1}^{#2}}
\nc{\G}[2]{G_{#1#2}}
\nc{\W}[2]{W_{#1#2}}
\nc{\D}[2]{D_{#1#2}}
\nc{\A}[2]{A_{#1#2}}
\nc{\Ap}[2]{A^\prime_{#1#2}}
\nc{\p}[2]{p^{#1}_{#2}}
\nc{\vv}[2]{v_{#1}^{#2}}
\nc{\rr}[2]{r_{#1}^{#2}}
\nc{\lij}[2]{l_{#1}^{#2}}

\nc{\Lsc}[2]{\scL_{#1#2}}
\nc{\Lrm}[2]{L_{#1#2}}
\nc{\Rsc}[2]{\scR_{#1#2}}

\nc{\spa}{SU(2)_1}
\nc{\spb}{SU(2)_2}
\nc{\spc}{SP(2 n - 4)}
\nc{\spd}{SP(4 n + 2 m - 10)}

\newcommand{\drawsquare}[2]{\hbox{%
\rule{#2pt}{#1pt}\hskip-#2pt
\rule{#1pt}{#2pt}\hskip-#1pt
\rule[#1pt]{#1pt}{#2pt}}\rule[#1pt]{#2pt}{#2pt}\hskip-#2pt
\rule{#2pt}{#1pt}}

\newcommand{\Yfund}{\raisebox{-.5pt}{\drawsquare{6.5}{0.4}}}
\newcommand{\Ysymm}{\raisebox{-.5pt}{\drawsquare{6.5}{0.4}}\hskip-0.4pt%
        \raisebox{-.5pt}{\drawsquare{6.5}{0.4}}}
\newcommand{\Yasymm}{\raisebox{-3.5pt}{\drawsquare{6.5}{0.4}}\hskip-6.9pt%
        \raisebox{3pt}{\drawsquare{6.5}{0.4}}}
\begin{document}

\begin{titlepage}

\begin{center}

\bigskip

\vspace{2cm}

{\hbox to\hsize{hep-th/9605113 \hfill EFI-96-15}}
{\hbox to\hsize{ \hfill Fermilab-Pub-96/107-T}}
{\hbox to\hsize{ \hfill  revised version, to appear in {\em Nucl. Phys.} {\bf B} }}

\bigskip

\bigskip

\bigskip

{\Large \bf Duality and Exact Results in Product Group Theories}

\bigskip

\bigskip

{\bf Erich Poppitz}$^{\bf a}$, {\bf Yael Shadmi}$^{\bf b, c}$
and  {\bf Sandip P. Trivedi}$^{\bf b, d}$ \\

\bigskip

\bigskip

$^{\bf a}${\small \it Enrico Fermi Institute\\
 University of Chicago\\
 5640 S. Ellis Avenue\\
 Chicago, IL 60637, USA\\

{\rm email}: epoppitz@yukawa.uchicago.edu\\}

\smallskip

 \bigskip

$^{\bf b}${ \small \it Fermi National Accelerator Laboratory\\
  P.O.Box 500, Batavia\\
  IL 60510, USA\\

$^{\bf c}${\rm email}: yael@fnth06.fnal.gov \\

$^{\bf d}${\rm email}: trivedi@fnal.gov\\ }

\vspace{1.3cm}

{\bf Abstract}

\end{center}

We study the non-perturbative behavior of some $N=1$
supersymmetric product-group gauge theories 
with the help of duality. As a test case we investigate an 
$SU(2) \times SU(2)$ 
theory in  detail. Various  dual theories are constructed  using known simple-group
duality for one group or both groups in succession. Several stringent
tests show that the low-energy behavior of the dual theories agrees with
that of the electric theory. When the theory is in the  confining phase 
we calculate the exact superpotential. Our results strongly
suggest that, in general,  dual theories for product groups can be constructed
in this manner, by  using simple-group duality  for both groups. Turning to
a class of theories with $SU(N) \times SU(M)$ gauge 
symmetry we study the renormalization group 
flows in the space of the two gauge couplings and show that they are
consistent with the absence of phase transitions. Finally, we show that a subset of these theories, with $SU(N) \times SU(N-1)$ symmetry break supersymmetry dynamically. 

\end{titlepage}

\renewcommand{\thepage}{\arabic{page}}
\setcounter{page}{1}

\baselineskip=18pt

\mysection{Introduction and Summary.}

\label{introductionandsummary}

\subsection{Motivation.}

\label{motivation}

The past two years have seen dramatic progress in the study of
the non-perturbative behavior of supersymmetric (SUSY) gauge theories. Duality,
which relates the strongly coupled behavior of one gauge theory to
the weakly coupled behavior of another,
has emerged as a key idea in this
study~\cite{seibergwitten},
\cite{seiberg}, \cite{DualityReview},
\cite{seibergexact}.
Most of the work in this context has focused on gauge theories
with simple gauge groups. While some work has been done involving theories
with non-simple groups \cite{ILStrassler},
one would like to understand them in more
detail. There are several reasons for this:
\begin{enumerate}
\item
Such an investigation will serve as a non-trivial check of simple-group
duality.  Gauging a global symmetry in two theories related by 
Seiberg duality is often a relevant perturbation,
and  the equivalence of the
resulting two theories will give further evidence for duality.
\item Product groups often arise in the course of dualizing theories with
simple groups, once one goes beyond the simplest matter
representations \cite{berkooz}, \cite{p1}, \cite{pouliotstrassler}, \cite{luty}.

\item  Many phenomenologically interesting chiral gauge theories 
consist of product gauge groups.
 
\item Several classes of product group theories exhibit dynamical SUSY
breaking.
\end{enumerate}

In  this
paper
we focus on a relatively simple product-group theory:
an $SU(2) \times SU(2)$ gauge theory. We  construct duals to this theory
and study its non-perturbative behavior. 
 We expect that the insights obtained are applicable 
to more complicated product-group
theories. In particular,  we  extend some  of our analysis  to 
 $SU(N) \times SU(M)$ theories.

Since we will closely follow Seiberg's original work  \cite{seiberg} it is
useful to
briefly review his main results here. Seiberg studied SUSY $SU(N)$ gauge
theories with $N_f$ flavors of fundamental matter fields. He found that
when $N_f \le N_c + 1$ the theory confines and its superpotential is determined
by holomorphy and symmetries. When $N_f > N_c + 1$,
there are  points in the  moduli space where extra particles become
light.
In this regime Seiberg constructed an $SU(N_f-N_c)$ theory and gave
strong arguments showing that it has the same low-energy behavior as the
$SU(N_c)$ theory. Moreover,  in a sense,  as the original theory becomes more
strongly coupled the $SU(N_f-N_c)$ theory becomes more weakly coupled.  Thus,
one could regard it as being dual to the $SU(N_c)$ theory.

\subsection{The ${\bf SU(2) \times SU(2)}$ Theories.}

In the first part of this paper,  (sections \ref{thenmmodels} -
\ref{theconfiningmodels}), we extend Seiberg's  results to the $SU(2)_1\times
S U(2)_2$ theory. The theory we study has
$2n$ $SU(2)_1$ fundamentals, $2m$ $SU(2)_2$ fundamentals, and one field 
transforming as a fundamental under both groups. We will refer to this theory as
the $[n,m]$ model. We will analyze the theory as $n$, $m$ are
varied\footnote{One simplifying feature of this theory is that it is
non-chiral.
The ability to add mass terms for all fields  provides
better control on its infra-red behavior.}.
As in the case of SUSY QCD, we will find that for small values of $n$ and $m$,
($n,m\le 2$), the theory is confining.  For larger values of $n$ and $m$
the theory can be in the non-Abelian Coulomb phase, and we construct dual
descriptions for it. The analysis in this case is qualitatively different
depending on whether $n,m > 2$, or only one of them is greater than 2.
 We discuss these different possibilities below.

\subsubsection{The Duality Regime.}

We begin our study of duality
in section~\ref{thenmmodels}
by considering the $[n,m]$ models with both $n,m \ge 3$.
In this case, each $SU(2)$, considered separately,
has $N_f > N_c + 1 = 3$ flavors, and
one expects
a dual theory to exist.
In fact,
with  some thought, several theories can be constructed which could,
potentially, have the same low-energy behavior as the original
$[n,m]$ theory (we will sometimes refer to this theory as  the electric theory).
For example, one can turn off, at first, the gauge coupling of the second
gauge group. The resulting $SU(2)$ gauge theory has a well known dual which
has a  global symmetry corresponding to the second $SU(2)$. It is natural
to guess that on gauging this symmetry one gets a theory which agrees with the
electric one in the infra-red. One can now carry this process one step further
and dualize the second $SU(2)$ symmetry as well, thereby getting another dual
theory.
Note  that by construction these theories have the same global symmetries as
the
original electric one, and  the 't Hooft anomaly matching
conditions for these symmetries are  satisfied.

Dualizing $SU(2)_1$ first,
 we construct two dual theories. One with gauge group $\spc\times\spb$,
and the other with gauge group 
$\spc\times \spd$\footnote{We use $SP(N)$
duals rather than $SU(N)$ duals, as the global symmetries are more manifest
in them.}.  Dualizing $SP(2)_2$ first, one would
obtain similar duals, with $n$ and $m$ exchanged.

The question we investigate is this: 
do these dual theories really have the same
infra-red physics as the original electric theory? To analyze this,
the low energy behavior  is probed in two   different ways:

First,
in section \ref{massflowsandscalematching},
mass terms are added for  some of the fields in the
electric theory. The electric theory then flows to a  new low-energy
theory.
As we show,  the dual theories flow to the duals of this  low-energy theory,
and the relations between
the strong coupling scales of the electric and dual theories change
 consistently  in the process.

Second,
in section  \ref{deformationsalongflatdirections},
 the moduli spaces of the electric and dual theories
are shown to agree by comparing  various flat directions in them.
We find that along these flat directions, the two 
theories are related by simple-group duality.
In particular, we establish that the chiral rings in the
two theories are the same.

Taken together, these  checks  strongly suggest
that the dual theories  have the same low energy behavior as the  electric
theory.
Thus, while these duals  arise naturally when one gauge coupling is much bigger
than the other,
 they are in fact valid for arbitrary ratios of the couplings.

\subsubsection{The ``Partially Confining'' Models.}

In section 3 we study the 
 ``partially confining'' models, in which one 
of the  groups, say $SU(2)_1$, is confining, when the other gauge
coupling  is turned off. This class includes the $[2,m]$ and $[1,m]$ models.
 A convenient starting point for studying the electric theory is
the limit $\Lambda_1 \gg \Lambda_2$, where $\Lambda_1$, $\Lambda_2$ are the 
strong coupling scales of $\spa$, $\spb$ respectively.
In this limit  the first gauge group  confines at the scale $\Lambda_1$, 
generating a non-perturbative superpotential.
Below this scale  one can use an effective theory in terms of the $SU(2)_1$ 
mesons, some of which transform under $SU(2)_2$. The low energy theory is therefore
an $SU(2)$ gauge
theory.

One can use $SP$ duality to construct an $SP(2m-2)$ dual of this theory.
It is interesting, however, to see whether the same theory is obtained by 
flowing down from the $[n,m]$ duals we discussed above.  
We analyze this question in  section~\ref{the2mmodels}.
The relevant dual to consider
 is the $[3,m]$ dual with gauge group $SU(2)\times SP(2m+2)$.
On flowing to the $[2,m]$ theory, the $SU(2) \times SP(2m+2)$ dual theory is 
indeed
broken to an $SP(2m-2)$ subgroup.
One can show that the non-perturbative
superpotential of the electric theory must  arise in the  dual  theory from
instanton-like  configurations with winding in both the $SU(2)$ and
the {\it partially broken} $SP(2m+2)$ subgroups.
This makes this case
somewhat different from the
non-perturbative effects in simple-group theories which  arise only when
the dual group is completely broken.
We expect the  non-perturbative
configurations in the present  case to include, but not be
restricted to instantons lying
in the diagonal $SU(2)$ subgroup of $SU(2) \times SP(2m+2)$.
With the non-perturbative superpotential
in place, this dual theory has the same  infra-red behavior as the electric
one.

We then turn to
the $[1,m]$ models
considered in section \ref{theflow2m1m}. 
Here the electric theory itself  is intriguing.
In the limit $ \Lambda_1 \gg\Lambda_2$
$SU(2)_1$ has a quantum modified moduli space.
  Symmetry considerations show that 
an axion-dilaton field must arise
 in the low-energy effective theory
in order to cancel  anomalies via the Green-Schwarz
mechanism.
However, since this field is generated dynamically and symmetries do not fix it
uniquely, it is not straightforward  to determine it. Duality provides
a convenient way to do so.  Starting with the  $SP(2m-2)$ dual theory
described above and  flowing  to the $[1,m]$  case, one
finds that the  theory is higgsed, as usual, but the scale at which
it is broken  depends on a modulus. This modulus is the required
dilaton in the  electric theory. With this identification, the electric theory
and the resulting dual once again agree in the infra-red.

There is another dual theory for the $[1,m]$ case obtained by flowing
down from the $[3,m]$ dual with gauge group
$SP(2m-4) \times SP(4m-4)$.
While we have not analyzed it in generality, we show in 
section \ref{the1mmodels} that for $m=3$,
this dual  too reproduces
the infra-red behavior of the electric theory.
This already is quite remarkable since the electric theory
has a quantum modified moduli space.

Note that while we analyze  the electric theory in the limit
$\Lambda_1 \gg \Lambda_2$, the dual descriptions we use are valid more
generally. Their equivalence  to the electric theory indicates
 that the electric description
too is valid for all values of $\Lambda_1/\Lambda_2$.

\subsubsection{Conclusions From the Study of the Duality Regime.}
 
The central lessons that emerge from this study of duality
 are as follows:

First, as mentioned in the very beginning, from the point of view of
duality in simple-group theory, gauging an additional group provides a highly
non-trivial consistency check of duality.

Second, we have established that dual theories can be constructed  
by alternately 
dualizing the various factors in a product group. This reduces the problem
of constructing dual theories for the product group cases to the simpler
problem of constructing duals for each of the individual groups.
The infra-red equivalence of the electric and dual theories follows from
simple-group duality in the limit when one gauge coupling is much
bigger than the other. But to establish this in general requires the
detailed tests described above.

Finally, all the  evidence obtained is consistent with  the behavior of the 
product-group
theory changing smoothly as the ratio of the two gauge couplings is varied.
In particular  there is no evidence for a phase transition, which would
drastically alter the infra-red behavior\footnote{
There is some lore to the effect that such transitions
cannot occur  in supersymmetric theories \cite{seibergwitten},
\cite{phasesofn1theories}.
Our results are in accord with this.}.

While we have only studied  in detail the $SU(2) \times SU(2)$ case,
we  expect most of our results to hold in general
for product-group theories.

\subsection{The Confining Models.}

We end our study of the $SU(2) \times SU(2)$ theories
by considering the confining models 
 in section \ref{theconfiningmodels}.
By analyzing the theory in the two limits, $\Lambda_1 \gg \Lambda_2$
and $\Lambda_2 \gg \Lambda_1$
we find the exact superpotentials and the massless particles in these models.
We show that the descriptions obtained in these two limits agree
and in fact argue that they  should be valid more generally, for all
values of $\Lambda_1/\Lambda_2$. Again, this is in agreement with the picture 
emerging from studying the dual models, as the confining models can be obtained by
 flowing down from the dual theories.

We derive the superpotentials of the confining models by  adding the
contributions generated by the two groups. We expect this to be true
quite generally as well.

Finally, we note that our discussion dealt with theories which were driven
into the confining regime by adding suitable mass terms.
One can also go into this  regime by adding Yukawa terms,
without accompanying mass terms
(we give an example of this in Sect. \ref{flowsbyyukawaperturbations}).
The exploration of such theories, especially from the point of view of
supersymmetry breaking, is left for  the future.

\subsection{The ${\bf SU(N) \times SU(M)}$ Theories and Supersymmetry
Breaking.}

We conclude
in section \ref{thesuntimessummodels}
by studying some features of the more general $SU(N) \times SU(M)$
theories.

First,
in section \ref{renormalizationgroupflows}
 we study  the renormalization group flows in the space
of the two
gauge couplings. We do this in the vicinity of the
two fixed points
obtained by  turning off one or the other gauge coupling. We find a simple
criterion to decide when the gauge coupling that  is initially turned off,
is a relevant perturbation. We then use this criterion to argue that the flows
are    consistent locally 
with the absence of a phase transition.

Second,
in section \ref{dualityinsuntimessum},
we extend our results for the $\spa\times\spb$ theories and construct a 
set of duals  for the $SU(N)\times SU(M)$ theories.

Finally,
in section \ref{suntimessun-1andsupersymmetrybreaking},
as an illustration of the richness of this class of theories, we
analyze a subset consisting of $SU(N) \times SU(N-1)$ theories
and show that they break supersymmetry after adding appropriate matter
fields and Yukawa couplings.

The presentation in the paper follows the order outlined in the 
introduction.


\mysection{The  ${\bf SU(2)_1 \times SU(2)_2}$  Models.}

\label{thenmmodels}

In this section we study the non-perturbative dynamics of an 
$SU(2)_1\times SU(2)_2$  gauge theory with matter fields in the fundamental representation. We explore the different phases of the theory as its matter content is varied. We note that the $SU(2)_1\times SU(2)_2$ theory is non-chiral  --  all its matter fields can be given mass terms. This allows for a variety of probes of the infrared physics.

The theory we consider has 
one field, $\Q\al\ald$,  transforming as a fundamental under both groups, $2n$ fields transforming as fundamentals of the first group, $\Lrm{\al}i$,  and $2m$ fields transforming as fundamentals of the second group, $\R{\ald}a$. Here $\al=1,2$, $\ald=1,2$ are the gauge indices of
$\spa$, $\spb$ respectively 
and $i=1\ldots 2n$, $a=1\ldots 2m$. We will refer to this theory as the $[n,m]$ theory, or the $[n,m]$ model. The field content of the $[n,m]$ theory is summarized in Table~1.

In analogy to QCD we sometimes refer to $SU$- and $SP$-fundamentals
as ``quarks'',
and to
a pair of fundamentals as one ``flavor''.
The $[n,m]$ theory has the non-anomalous global symmetry group
$SU(2n)\times
SU(2m)\times U(1) \times U(1)_R$. For $n,m \ge 0$ there is a one-parameter
family of $U(1)_R$ symmetries. As a result,  $U(1)_R$ charges of fields and
their dimensions at superconformal infra-red fixed points are not uniquely
determined.
The charges of
the fields under the nonanomalous global symmetries are also given
in Table 1.
Our conventions and notations are summarized in Appendix A.
We denote the strong coupling scales of the two factors in
$SU(2)_1 \times SU(2)_2$
by $\Lambda_1$ and $\Lambda_2$ respectively. The charges of
the fields and the
strong coupling scales under the various anomalous symmetries
are given in Appendix B.
%
%
\begin{table}\begin{center}
{\centerline {Table~1: Field Content of Electric Theory}}
\vspace{0.2cm}
\label{tab1}

\begin{tabular}{c|c|c|c|c|c|c} \hline\hline
$\ $ & $SU(2)_1$ & $SU(2)_2$ & $SU(2n)$ & $SU(2m)$ & $U(1)$ &
$U(1)_R$ \\
\hline
 & & & & & &\\
%
$Q_{\al\ald}$ & $\Yfund$ & $\Yfund$ & 1   & 1   & $-mn$ & $m-1$   \\
%
$L_{\al i}$   & $\Yfund$ & 1   & $\Yfund$ & 1   & $m$   & $1-m/n$  \\
%
$R_{\ald a}$  & 1   & $\Yfund$ &  1  & $\Yfund$ & $n$   & 0        \\
 \hline\hline
\end{tabular}
\end{center}
\end{table}
%

\subsection{The Duals of the ${\bf [n,m]}$ Models.}

\label{thedualsofthenmmodels}

In this section we will construct theories dual to the $[n,m]$ models.
Our basic building block will be the dual of an $SP(N)$ gauge theory
with fundamental matter first constructed in \cite{seiberg} and
subsequently studied in \cite{sp}. By applying this $SP$ duality to one or
both of the $SU(2)$ groups we will construct
two kinds of dual theories\footnote{Other dual
theories can also be constructed involving $SU(N)$
groups. We focus on the $SP$ duals here  since the global symmetries are more
manifest in them. }. The dual theories, by construction,  will
have the same global symmetries as the original electric theory and
the 't Hooft anomalies for these symmetries will match with those
in the   electric theory.  In the following sections we will subject the dual
theories to
other non-trivial checks of duality.
In Section \ref{massflowsandscalematching} we will change the infra-red
behavior
of the electric theory by adding mass terms and show that the dual theories
flow to new ones in a consistent way. In Section
\ref{deformationsalongflatdirections} we will  study
the consistency of duality with deformations along flat directions.
This test is crucial
in verifying that the moduli spaces and chiral rings of the electric
and magnetic theories
are the same. Some relevant information regarding duality
in $SP$ groups is summarized in Appendix A.

As mentioned above, two kinds of theories  dual to the $[n,m]$ models
will be constructed.
We discuss them in turn below. In constructing the
first dual to the $SU(2) \times SU(2)$ theory
it is useful to  consider the
theory in the limit in
which the $\spb$ strong coupling scale, $\Lambda_2$, 
is negligible compared to the $\spa$
scale, $\Lambda_1$.
In this limit one has an $\spa$ theory
with $n+1$ flavors. For $n \ge 3 $, it has an equivalent 
 infra-red description
in
terms of a dual theory with gauge group $\spc$
\cite{sp}. The dual theory has $n+1$ flavors of ``dual quarks'',
 $q_\lam^{\ald}$
and
$\lij{\lam}i$, transforming as fundamentals of $\spc$, and
$\lam=1... 2n-4$\ is
the $\spc$ gauge index. In addition, the theory contains
$\spc$ singlet fields
which are mapped to the mesons of the original $\spa$:
\beq
\label{su21mesons}
X  \equiv {1 \over 2} \varepsilon^{{\al_1}{\al_2}}
\varepsilon^{{\ald_1}{\ald_2}}
Q_{{\al_1}{\ald_1}} Q_{{\al_2}{\ald_2}}~ , ~ ~ ~ ~
 \Lsc{i}{j}\equiv  L_i \cdot L_j ~ ,  ~ ~ ~ ~ \V{\ald}i
 \equiv Q_{\ald}\cdot L_i ~ ,
\eeq
 where
the product denotes $\spa$ contraction (our conventions are also
given in Appendix A).
The dual theory has a superpotential~\cite{sp}:
\beq
\label{dual1spotential}
W\, =\, {1\over 4\mu_1} ~\left( -X\, q^{\ald_1} \cdot q^{\ald_2}
\varepsilon_{\ald_1\ald_2} \, +\,
2~\V{\ald}{i}\,  {q}^{\ald}\cdot l^i \, +\,
\Lsc{i}{j}\, l^i\cdot l^j \right) ~.
\eeq
The dimension-one parameter  $\mu_1$
is introduced in order to match the dimensions of the electric and magnetic
mesons in the ultraviolet \cite{seiberg}.
The parameter $\mu_1$ and the strong coupling scales
$\Lambda_1$
of $\spa$, and  $\Lambda^\prime_1$ of $\spc$ satisfy the scale
matching relation
(\ref{spscalematching}) \cite{sp}:
\beq
\label{match1}
\Lambda_1^{5-n}{ \Lambda^\prime}_1^{2n-4}\, = \, 16\,(-1)^{n-1}\,
\mu_1^{n+1} ~.
\eeq

We note in particular, that this $SP(2n-4)$ theory has a global $SU(2)$
symmetry corresponding to $SU(2)_2$ in the electric theory. By gauging it
in the $SP(2n-4)$ theory one expects to get a dual to the $SU(2)_1 \times
SU(2)_2$ theory. The resulting $\spc \times SU(2)_{2^\prime}$ theory
will be referred to as the {\bf first dual}.
Its field content is
summarized in Table~2 (in order to avoid confusion we refer to the
$SU(2)_2$ symmetry in the dual as $SU(2)_{2^\prime}$).
 
\begin{table}\begin{center}
{\centerline {Table~2: Field Content of First Dual}}
\vspace{0.2cm}
\label{tab2}

\begin{tabular}{c|c|c|c|c|c|c} \hline\hline
$\ $ & $SP(2n-4)$ & $SU(2)_{2^\prime}$ & $SU(2n)$ & $SU(2m)$ & $U(1)$ &
$U(1)_R$ \\
\hline
 & & & & & &\\
$q_{\lam}^{\ald}$ & $\Yfund$ & $\Yfund$ & 1   & 1   & $mn$ & $2-m$   \\
$l_\lam^i$   & $\Yfund$ & 1   & ${\overline{\Yfund}}$ & 1   & $-m$   & $m/n$ \\
$R_{\ald a}$  & 1   & $\Yfund$ &  1  & $\Yfund$ & $n$   & 0        \\
$V_{\ald i} $ & 1 & $\Yfund$ & $\Yfund$ & 1 & $m-mn$ & $m-m/n$ \\
$X $ & 1 & 1 & 1 & 1 & $-2mn$ & $2(m-1)$ \\
$\scL_{ij} $ & 1 & 1 & $\Yasymm$ & 1 & $2m$ & $2(1-m/n)$ \\
 \hline\hline
\end{tabular}
\end{center}
\end{table}
%
%

The $SU(2)_{2^\prime}$ gauge group has $2 ( 2 n + m - 2)$
fundamentals\footnote{We use the
 antisymmetric tensor to lower the
$SU(2)_{2^\prime}$ index of $q$, to conform
with our definition of $SP$ doublets as having lower gauge indices.}:
\beq
\label{su2primeflavors}
 \varepsilon_{\ald\ald_1} {q}_\lam^{\ald_1}
\ ,
 {1\over \mu_1} \V{\ald}i \ , \ \R{\ald}a \ .
\eeq
The Wilsonian gauge coupling (strong coupling scale)
of $SU(2)_{2^\prime}$ is uniquely determined,  by
its charges under the various anomalous symmetries, given in Table 6
of Appendix B, to be:
\beq
\label{lambda2prime}
{\Lambda_2^\prime}^{8-2n-m}\, =  \,
(-1)^n\, 2^{n-2}\,
{ \Lambda_1^{5-n} \Lambda_2^{5-m} \over
  \mu_1^{2+n} }~ .
\eeq
The numerical coefficient above is derived at the end of this section
 by requiring consistency of this
dual with deformations along the $X \ne 0 $ flat
direction.

Although the construction of the $SP(2n-4) \times SU(2)_{2^\prime}$
dual above was motivated by
considering the electric theory in the limit $ \Lambda_1 \gg \Lambda_2$, 
we will find through several checks in the subsequent sections that the
electric and dual theory  in fact agree  in the infra-red regardless
of the value of $\Lambda_1/\Lambda_2$. 
As a first test we note here that the
't Hooft anomalies in the electric theory and the dual are guaranteed to
match by construction (this can also be checked explicitly by
using Tables 1 and 2). It is also worth mentioning that one can
clearly repeat the above mentioned procedure to
dualize $SU(2)_2$ instead of $SU(2)_1$ thereby obtaining  an
$SU(2)_{1^\prime} \times SP(2m-4)$ dual theory. The analysis of this dual
is very analogous to that of the $SP(2n-4) \times SU(2)_{2^\prime}$ theory;
consequently we focus on the latter in this paper.

We now extend this process one step further
by dualizing $SU(2)_{2^\prime}$ in the first dual thereby
giving another  dual theory which we will call
the {\bf second dual}.
The dual of $SU(2)_{2^\prime}$ with $2 n + m - 2$ flavors
(\ref{su2primeflavors})
is an
$SP(4n+2m-10)$  gauge theory  with $2 n + m - 2$ flavors
$\p\lam\lamd$, $\vv{\lamd}i$ and $\rr{\lamd}a$, where
$\lamd=1\ldots 4n+2m-10$
is the $SP(4n+2m-10)$ gauge index.
In addition, the $SU(2)_{2^\prime}$   mesons constructed from
(\ref{su2primeflavors})
appear as basic fields:
\beq
\label{sp2primemesons}
\begin{array}{lll}
\A{\lam_1}{\lam_2} \equiv  q_{\lam_2} \cdot  q_{\lam_1}~, &
{}~ ~  \D{\lam}i \equiv  q_\lam \cdot  V_i ~,  &
{}~ ~ \G{\lam}a \equiv q_\lam \cdot R_a ~, \\
 \W{i}{j} \equiv V_i\cdot V_j ~,   &  ~ ~
 \Y{i}{a} \equiv \V_i \cdot  R_a ~,
& ~ ~ \scR_{ab} \equiv  R_a \cdot R_b ~,
\end{array}
\eeq
where for conciseness we have omitted in eq. (\ref{sp2primemesons}) 
the various powers of $\mu_1$ from
(\ref{su2primeflavors}) that 
are needed to correctly match the dimensions.
The full gauge symmetry in the theory is then
$\spc\times\spd$.
The theory  has the superpotential:
\beqa
\label{intermediatespotential}
W\, &=&\, {1\over 4\mu_1}
 \left( X~ \A{\lam_1}{\lam_2}\, J^{\lam_1\lam_2} -
2 ~\D{\lam_1}i\, \lij{\lam_2}i ~J^{\lam_1\lam_2} +
\Lsc{i}{j}\, \lij{\lam_1}i ~ \lij{\lam2}j ~J^{\lam_1\lam_2}
\right) \nonumber \\
&+& {1\over 4\mu_2}  \left( \Rsc{a}{b}\, r^a\cdot r^b +
2~\G{\lam}a\,  p^\lam\cdot r^a + {2\over \mu_1}~\Y{i}{a}\, v^i\cdot r^a
\right. \\
& ~ & \left. \hspace{1.5cm} + ~
\A{\lam_1}{\lam_2}\, p^{\lam_1}\cdot p^{\lam_2}  +
{2\over \mu_1}~\D{\lam}i\, p^\lam\cdot v^i  +
{1\over \mu_1^2}~\W{i}{j}\, v^i\cdot v^j \right)
\nonumber ~,
\eeqa
where the first three terms come from the
superpotential (\ref{dual1spotential}) expressed in terms of the fields of the
second dual, and $\mu_2$ is a dimension-one parameter needed to relate the   ultraviolet dimensions
of the mesons (\ref{sp2primemesons}) in the electric and magnetic theories.
As in eq. (\ref{match1}), the parameter $\mu_2$ and
 the scales $\Lambda_2^\prime$  of $SU(2)_{2^\prime}$ and
${\bar\Lambda}_2$ of $\spd$ obey the scale matching
relation (\ref{spscalematching}):
\beq
\label{match2}
{\Lambda_2^\prime}^{8-2n-m} {\bar\Lambda}_2^{4n+2m-10} \, =\,
16\, (-1)^m\,
\mu_2^{2n+m-2} \ .
\eeq
Substituting ${\Lambda_2^\prime}^{8-2n-m}$ from~(\ref{lambda2prime}) we find
the scale matching relation for the scale ${\bar\Lambda}_2$ of $\spd$:
\beq
\label{lambdab2}
 {\bar\Lambda}_2^{4n+2m-10} \, =\,
2^{6-n} \, (-1)^{n+m} \,
{ \mu_2^{2n+m-2} \mu_1^{2+n}\over
\Lambda_1^{5-n} \Lambda_2^{5-m} } \ .
\eeq

It follows from  the superpotential~(\ref{intermediatespotential}) that
the fields $X$, $\A{\lam_1}{\lam_2} J^{\lam_1\lam_2}$, $\D{\lam}i$ and
 $\lij{\lam}i$ are heavy in the second dual. Their equations of motion are:
\beqa
\label{eom}
\A{\lam_1}{\lam_2} ~ J^{\lam_2\lam_1} &=& 0 \, , \nonumber \\
X &=& {1\over 2n-4}\,
 {\mu_1\over \mu_2} \, J_{\lam_1 \lam_2} ~p^{\lam_1} \cdot p^{\lam_2} \, ,\\
\lij{\lam_1}i &=& \, {1\over \mu_2}\, J_{\lam_1 \lam_2} ~p^{\lam_2}\cdot v^i
\nonumber \, ,\\
\D{\lam}i &=& \, -\, \Lsc{i}{j} ~\lij{\lam}j \ . \nonumber
\eeqa
The first equation in~(\ref{eom})
 sets the trace part of the anti-symmetric field
$\A{\lam_1}{\lam_2}$ to zero. The remaining light fields (see Table~3) are
therefore the traceless part of the anti-symmetric field,
$\Ap{\lam_1}{\lam_2}$, which transforms under $\spc$ only, $\p\lam\lamd$, which
is a fundamental under both groups,
$\vv{\lamd}i$ and  $\rr{\lam}a$, which are $\spd$ fundamentals,
and $\G{\lam}a$, which are $\spc$ fundamentals.
The theory also contains
a number of singlets under both groups:
 $\Lsc{i}{j}$,   $\Rsc{a}{b}$, $\Y{i}{a}$ and  $\W{i}{j}$.
Together,
 these  fields saturate the `t~Hooft anomalies of the original $[n,m]$ theory.

%
\begin{table}\begin{center}
{\centerline {Table~3: Field Content of Second Dual}}
\vspace{0.2cm}
\label{tab3}

\begin{tabular}{c|c|c|c|c|c|c} \hline\hline
  & $SP(2n-4)$ & $SP( 4 n + $ & $SU(2n)$ & $SU(2m)$ & $U(1)$ &
$U(1)_R$ \\
  &         & $ ~ ~ 2 m - 10)$   &          &        &                 &
  \\
\hline
 & & & & & &\\
$p^{\lam}_{\lamd}$ & $\Yfund$ & $\Yfund$ & 1   & 1   & $-mn$ & $m-1$   \\
$v_{\lamd}^i$   & 1  & $\Yfund$ & ${\overline {\Yfund}}$ & 1   & $mn-m$   &
$1-m+{m\over n}$  \\
$r_{\lamd}^a$  & 1   & $\Yfund$ &  1  & ${\overline {\Yfund}}$ & $-n$   & 1
   \\
$\Ap{\lam_1}{\lam_2}$ & $\Yasymm$ & 1 & 1 & 1 & $2mn$ & $2(2-m)$ \\
$\G{\lam}a$ & $\Yfund$ & 1 &
1 & $\Yfund$ & $mn+n$ & $2-m$ \\
$\scL_{ij} $ & 1 & 1 & $\Yasymm$ & 1 & $2m$ & $2(1-{m\over
n})$ \\
$\Rsc{a}{b}$ & 1 & 1 & 1 & $\Yasymm$ & $2n$ & 0 \\
$Y_{i a}$ & 1 & 1 & $\Yfund$ & $\Yfund$ & $m+n-mn$ & $
m-{m\over n}$ \\
$\W{i}{j}$ & 1 & 1 & $\Yasymm$ & 1 & $2m-2mn$ & $2m-2{m\over
n}$ \\
 \hline\hline
\end{tabular}
\end{center}
\end{table}
%

To summarize, the second dual has an $\spc\times\spd$ gauge group with
an antisymmetric tensor and $ 2 n + 2 m - 5$ flavors of $SP( 2 n - 4)$,
$2 n - 2 + m$ flavors of $SP( 4 n + 2 m - 10)$, and has a superpotential:
\beqa
\label{finalspotential}
W &=& -\, {1\over 4\mu_1\mu_2^2}\, \Lsc{i}{j}\, v^i\cdot p^{\lam_1}
J_{\lam_1\lam_2}
v^j\cdot p^{\lam_2}\, +
\\
&+& {1\over 4\mu_2}\, (
\Rsc{a}{b}\, r^a\cdot r^b +
2\G{\lam}a\,  p^\lam\cdot r^a + {2\over\mu_1}\Y{i}{a}\, v^i\cdot r^a
+
\Ap{\lam_1}{\lam_2}\, p^{\lam_1}\cdot p^{\lam_2}  +
{1\over \mu_1^2}\W{i}{j}\, v^i\cdot v^j) \ .
\nonumber
\eeqa
 The scale ${\bar \Lambda}_1$ of
$SP(2n-4)$ in the second dual can be found by using
the various anomalous symmetries of Table 6 (Appendix B)
\beq
\label{lambdab1}
{\bar \Lambda}_1^{5-2m} =
 c(n, m)~{\Lambda_2^{5-m} \over \mu_1 ~\mu_2^{m-1} }\, ,
\eeq
up to a constant. We will show in Section \ref{massflowsintheseconddual}
that consistency of duality
with the mass flows implies recursion relations on $c(m,n)$ as $m$ and $n$
are varied, eqs. (\ref{cmrecursion}) and (\ref{cmrecursion2}).
Furthermore, in Section \ref{the1mmodels},
consistency of duality in
the  $[n,1]$ models will  fix $c(3,1) = - 2$, eq.  (\ref{cof1fixed}).
Together with the recursion relations
(\ref{cmrecursion}) and (\ref{cmrecursion2}) this will allow us to
determine the constant in eq. (\ref{lambdab1}):
\beq
\label{cmn}
c(n, m) = (-)^m~ 2^{ m + n  - 3} ~.
\eeq

Symmetry considerations do allow some field dependence in eq.
(\ref{lambdab1}), since there is one combination of fields and scales that
is invariant under all global symmetries.
 However, such a field dependence cannot occur. For example, it would introduce
unphysical singularities in the Wilsonian gauge coupling. Also,
it would not be consistent with the mass flows considered in the next section.

We should note that by repeating the procedure described
above in the opposite order and dualizing $SU(2)_2$ first followed
by $SU(2)_1$ we would obtain another dual theory of the second kind with
gauge group $SP(2n+4m-10) \times SP(2m-4)$. Additional duals
could  in principle be obtained   by continuing 
to  alternatingly dualize the two groups.
However,  $\spc$ now contains an antisymmetric tensor,
and its dual is still unknown.

We complete the construction of the $[n,m]$  duals  by
determining the constant in the scale matching relation 
(\ref{lambda2prime}) for
$\Lambda_2^\prime$  --  the scale
of $SU(2)_{2^\prime}$ in the first dual.
As was mentioned in the discussion preceding eq.(\ref{lambda2prime})
symmetries determine that
\beq
\label{lambda2primeNEW}
{\Lambda_2^\prime}^{8-2n-m}\, =  \,
 C \ \
{ \Lambda_1^{5-n} \Lambda_2^{5-m} \over
  \mu_1^{2+n} }~.
\eeq
We find the constant $C$ below by
going along an $X$ flat direction
and demanding consistency of this deformation with the first dual.
Along the $X \ne 0$ flat direction,
 the electric theory breaks to the diagonal
$SU(2)_D$ with the $2n+2m$ doublets $L_i$ and $R_a$, with scale
$\Lambda_D^{6-n-m}=\Lambda_1^{5-n}\Lambda_2^{5-m}/X^2$. In the
first dual (\ref{dual1spotential}), one flavor of
the dual quarks ($q$) becomes heavy and can be
integrated out.
The $SP(2 n - 4)$ theory then has $n$ flavors and confines,
thereby
generating a nonperturbative superpotential (\ref{spsuperpotential}).
Adding this superpotential to (\ref{dual1spotential}) (with the
fields $q$ integrated out and the result rewritten in terms of the
$SP(2 n - 4)$ mesons), it is easy to see that the only fields that remain light
are the
$SU(2)_{2^\prime}$-quarks $V_{\ald i}/\mu_1$, and that the
superpotential for the light fields vanishes.
The scale of the $SU(2)_{2^\prime}$ after integrating out the heavy fields
will be denoted by $\Lambda^\prime_{2L}$. Requiring that this scale
coincide with the
scale of the diagonal $SU(2)_D$ in the electric theory we find:
\beq
\label{xflatmatch}
{\Lambda^\prime_{2 L}}^{6-n-m}\ =\
 \left( {- X \over 2\mu_1} \right)^{n-2}\, C \,~
 { \Lambda_1^{5-n} ~ \Lambda_2^{5-m} \over  \mu_1^{n+2} }\  =\
 { \Lambda_1^{5-n}~ \Lambda_2^{5-m} \over X^2 } \ ,
\eeq
where the $ X  $ dependence of the middle term arises
since we integrate  out the fields $q$ at the mass scale 
$ - X/ 2\mu_1$, and use the scale matching relation
(\ref{scalematchingmass}).
To complete the identification of
$SU(2)^\prime$ with $SU(2)_D$ we further note that
the fields $V_i/\mu_1$  should be identified with $L_i$. This fixes
$\mu_1={\sqrt{X}}$. Solving for $C$ in eq. (\ref{xflatmatch}) now
we find that
\beq
\label{fixc}
C\, =\, (-1)^n\, 2^{n-2}\ ,
\eeq
thereby obtaining  the  coefficient in eq. (\ref{lambda2prime}).

\subsection{Mass Flows and Scale Matching.}

\label{massflowsandscalematching}

In this section we subject the above proposed duality to a strict
consistency check. First we change the infra-red behavior of the
electric theory by adding mass terms for some of its  fields. We then show that
the first and second  dual theories described above flow in the infra-red to
the appropriate duals of the low-energy electric theory. This constitutes
a non-trivial check on the equivalence of the low energy behavior of
the electric and dual theories. In the subsequent discussion we refer to the
renormalization group flow from the original infra-red theory to the new one,
obtained  after adding mass terms, as a ``mass flow''.

There are three basic mass flows to consider. First, upon giving a mass to one
flavor of $\spa$, such as $L_{i=1,2}$, the $[n,m]$ theory flows to the
$[n-1,m]$ theory. Similarly, giving a mass to one flavor of $\spb$, such as
$R_{a=1,2}$, one obtains the $[n,m-1]$ model. Finally, one can give a mass
to the field $Q$, which transforms under both groups,
  and  integrate it  out.
The resulting low energy electric theory then has two decoupled
$SU(2)$ gauge groups.
In the following we discuss these three different flows, first in the first
dual (Section \ref{massflowsinthefirstdual}),
and then in the second dual (Section \ref{massflowsintheseconddual}).

We will also demonstrate the consistency of the
scale matching relations
(\ref{match1}), (\ref{lambda2prime}), (\ref{lambdab2}) and (\ref{lambdab1})
 with the mass flows.
The meaning of the
scale matching relations is that the parameters $\mu_{1,2}$, the scales
$\Lambda_{1,2}$
of the electric theory and the scales $\bar{\Lambda}_{1,2}$ of the dual
theory have to obey
 (\ref{lambdab2}), (\ref{lambdab1}),
in order for the correlation functions of the electric and
magnetic theories to agree,  including
normalization. Similarly, in the first dual,
 $\Lambda_1^\prime$,
$\Lambda_2^\prime$ and $\mu_1$ have to obey (\ref{match1}),
(\ref{lambda2prime}). For any given value of $[n,m]$ the
parameters $\Lambda, \bar{\Lambda}$ and $\mu$ in the scale matching
relations can be
absorbed by redefining the various operators. It may thus seem that 
 the scale
matching relations are trivial. It is nontrivial,
however (see ref. \cite{david}), that the
scale matching relations, as we show below,
are  consistent with the various flows. In Section
\ref{thepartiallyconfiningmodels}, we
 provide additional nontrivial  checks
 on the scale matching relations when we consider the
properties of the partially confining models.

\subsubsection{Mass Flows in the First Dual.}

\label{massflowsinthefirstdual}

Since the first dual was obtained by dualizing $\spa$ only, it is not
affected
by the $[n,m-1]$ flow\footnote{Except for the scale $\Lambda_2^\prime$;
its change
is given by $m\rightarrow m-1$ in eq. (\ref{lambda2prime}).}.
We therefore turn to the $[n-1,m]$ flow.
 Adding a mass term
$M \Lsc{1}{2}$  for one flavor of $\spa$ in the electric theory, the
theory
flows to the
$[n-1,m]$ model, with the scale of $\spa$ given by
$\Lambda_{1 L}^{5-(n-1)} = M\, \Lambda_1^{5-n}$.

The first dual is higgsed to $SP(2n-6)\times SU(2)_{2^\prime}$ much
like in the case of a
single $SP$
group. We therefore do not discuss it further except to note the
change in the
$SU(2)_{2^\prime}$ scale. Adding the term $M \Lsc{1}{2}$ to the
superpotential~(\ref{dual1spotential}), the fields $l_1$ and $l_2$
get vevs
with $l_1\cdot l_2 = -2\, M\, \mu_1 \equiv v^2$.  
Plugging their  $D$-flat
vevs
\beq
\nonumber
l_\lambda^i= \left\{ \begin{array}{ll}
\delta_\lambda^i ~\sqrt{ - 2 \mu_1 M} &
\mbox{ if $\lambda = 1,2$} \\
0 & \mbox{ otherwise,~~}\end{array} \right.
\eeq
into the superpotential (\ref{dual1spotential}),
the $SU(2)_2$ doublets
$q_{1\ald}$, $q_{2\ald}$, $V_{{\ald}1}/\mu_1$ and $V_{{\ald}2}/\mu_1$ become
heavy, with mass $v/2$.
  $SU(2)_2$ now has $2(n-1)+m-2$ flavors, and its scale is, using
  (\ref{scalematchingmass}):
\beq
\label{lambdatwoprime}
{\Lambda_{2\, L}^\prime}^{8-2(n-1)-m} \, = \,
{v^2\over 4}\,
{\Lambda_{2}^\prime}^{8-2n-m} \, =\,
(-1)^{n-1} \, 2^{n-3}\,
{  {{\Lambda_1}_L}^{5-(n-1)}\, \Lambda_2^{5-m} \over \mu_1^{2+(n-1)} }\ ,
\eeq
in agreement with equation~(\ref{lambda2prime}) written in terms of
$\Lambda_{1_L}, \Lambda_{2_L}^\prime$.

The last mass flow to consider in the first dual is the decoupling of
 the common field $Q$.
Upon adding a mass term $M\, X$,
the electric theory consists of two disjoint $SU(2)$ groups, with $n$, $m$
flavors respectively, and scales
${{\Lambda_1}_L}^{6-n} = M\, \Lambda_1^{5-n}$ and
${{\Lambda_2}_L}^{6-m} = M\, \Lambda_1^{5-m}$.
Correspondingly, in the first dual, upon adding the term  $M\, X$ to the
superpotential
(\ref{dual1spotential})
the fields $q$ acquire vevs:
\beq
\label{qvevs1stdual}
q_\lambda^{\dot{\alpha}} = \left\{ \begin{array}{ll}
\delta_\lambda^{\dot{\alpha}} ~\sqrt{ - 2 \mu_1 M} &
\mbox{ if $\lambda = 1,2$} \\
0 & \mbox{ otherwise,~~}\end{array} \right.
\eeq
such that $SP(2n-4)\times SU(2)_{2^\prime}$ breaks to  $SP(2n-6) \times
SU(2)_D$,
where
$SU(2)_D$ is the diagonal group.
Plugging the vevs (\ref{qvevs1stdual}) into (\ref{dual1spotential}),
we see
that $V_{\ald i}$,  $l^i_1$ and $l^i_2$ are massive.
Their  equations of motion set them to zero
and the superpotential reduces to
\beq
\label{xspotential1}
W\ = \ {1\over 4\mu_1} ~\Lsc{i}{j} ~ l^i_{\lambda_1} ~ l^{j}_{\lambda_2}
{}~ J^{\lambda_1 \lambda_2} \ ,
\eeq
with $\lambda_{1,2} = 3 \ldots 2 n - 4$.
The light fields include the singlets $\Lsc{i}{j}$, the $SP(2n-6)$
fundamentals
$l^{i=1..2n}$, and the $SU(2)_D$ doublets $R_{a=1..2m}$.
Clearly, $SP(2n-6)$ is dual to the electric $SU(2)_1$,
as is evident from applying (\ref{scalematchingflat}):
$\Lambda_{1_L}^{\prime ~ 2 n - 6} = { 2 \over - 2 M \mu_1 }
\Lambda_1^{\prime ~ 2 n - 4}$,
which shows that $\Lambda_{1_L}^\prime$ is
precisely the scale of the dual of the low-energy electric theory with
${{\Lambda_1}_L}^{6-n} = M\, \Lambda_1^{5-n}$.
The other unbroken group in the low-energy theory,  $SU(2)_D$,
 should be identified with the
electric $SU(2)_2$, which was left untouched by the first duality
operation.
Matching the scale of the diagonal $SU(2)_D$ to the scales of
$SP(2n-4)$ and $SU(2)_{2^\prime}$, and using~(\ref{match1}),
(\ref{lambda2prime}),  we obtain
$
\Lambda_D^{6-m} \sim  \langle q^1\cdot q^2\rangle
{\Lambda_1^\prime}^{2n-4} {\Lambda_2^\prime}^{8-2n-m}  \sim
 M  \Lambda_2^{5-m} $
so that the scale of the diagonal $SU(2)_D$ is equal to the scale
of the original
$SU(2)_2$ as expected\footnote{The numerical factor in the relation between
$\Lambda_D$, $\Lambda_1^\prime$ and $\Lambda_2^\prime$ arises because we are
working in the ${\overline{\rm DR}}$ scheme here.}.

\subsubsection{Mass Flows in the Second Dual.}

\label{massflowsintheseconddual}

The flow $[n,m]\rightarrow [n,m-1]$ is the simplest flow in the second dual,
since $\spb$ was dualized last.
Adding a mass $M$ for the first flavor of $\spb$ in the electric theory
corresponds to adding the term
$M\Rsc{1}{2}$ to the  superpotential~(\ref{finalspotential}) in the second
dual.
The scale of the low-energy electric $SU(2)_2$ is $\Lambda_{2 ~L}^{5 - (m - 1)}
 = M \Lambda_{2 }^{5 - m}$ (\ref{scalematchingmass}).
In the dual theory, the  equation of motion for $\Rsc{1}{2}$ is
\beq
\label{r12eom}
r^1\cdot r^2 = v^2 = -2 \,M \mu_2
\eeq
Taking
the vevs of $r^a$ along the D-flat directions
\beq
\label{rvevs2nddual}
r_{\lamd}^a = \left\{ \begin{array}{ll}
\delta_{\lamd}^a ~{\sqrt{-2 M \mu_2}} &
\mbox{ if ${\lamd}= 1,2$} \\
0 & \mbox{ otherwise,~~}\end{array} \right.
\eeq
 we find that $SP(4n+2m-10)$ is higgsed down to
$SP(4n+2(m-1)-10)$ and that the superpotential~(\ref{finalspotential}) gives
mass to the  $\lamd = 1,2$
 components of all  $SP(4n+2m-10)$ fundamentals as well as to
the singlets $\Rsc{1}{2}$, $\Rsc{1}{a}$,  $\Rsc{2}{a}$, $\G{\lam}{1}$,
$\G{\lam}{2}$, $\Y{i}{1}$ and  $\Y{i}{2}$. The resulting low-energy
theory is therefore
$SP(2n-4) \times SP(4n+2(m-1)-10)$ and its superpotential is as
in~(\ref{finalspotential}) with $a$ now taking $2(m-1)$ values and
$\lamd = 3 \ldots 4 n+2 m -10$, as can be seen by integrating out
the heavy fields.

Note also that $SP(2n-4)$ now has two fewer flavors, since the fields
$\p{\lam}{\lamd}$ with $\lamd = 1,2$, and the
fields ${1\over \mu_2} G_{{\lam} a = 1,2}$ become heavy, with mass $v/2$.
It therefore has $2n+2(m-1)-5$ flavors
as expected in
$[n,m-1]$ case. The scale $\bar{\Lambda}_{1 L}$ of the low-energy
$SP(2 n -4)$ theory can be found by matching at the scale of the
mass of the heavy flavors (\ref{scalematchingmass}):
\beqa
\label{matchingmflow1}
\bar{\Lambda}_{1 L}^{5 - 2(m-1)}
&=& - { M \mu_2 \over 2 } ~ c(n, m) ~
 { \Lambda_2^{5 - m} \over \mu_1 ~\mu_2^{m - 1} } =
 - {1 \over 2} ~ c(n, m) ~{ \Lambda_{2 L}^{5 - (m-1)} \over \mu_1 ~
\mu_2^{(m-1)-1} } \\
&=&
c(n, m-1) ~ {\Lambda_{2 L}^{5 - (m-1)} \over \mu_1 ~\mu_2^{(m-1)-1} }~
,\nonumber
\eeqa
where the equality on the second line above is (\ref{lambdab1}) with $m
\rightarrow m-1$.
 Consistency of
the $[n,m] \rightarrow [n,m-1]$
 flow with  (\ref{lambdab1}) therefore implies a recursion relation for $c(n,
m)$:
\beq
\label{cmrecursion}
c(n, m) = - 2 ~c(n, m-1)~.
\eeq

Finally, the scale of $SP(4 n + 2 (m -1) -10)$ in the low-energy  magnetic
theory is
(\ref{scalematchingflat}):
\beq
\label{flowlamb2p}
\bar{\Lambda}_{2 L}^{4n+2(m-1)-10} \, =\,{ 2\over\langle r^1\cdot r^2\rangle }~
{\bar\Lambda}_2^{4n+2m-10}
\, =\, - \,
{{\bar\Lambda}_2^{4n+2m-10}  \over
M\mu_2 } \ .
\eeq
This is clearly consistent with  (\ref{lambdab2}) after taking
$m\rightarrow m-1$ in it and identifying
$\Lambda_{2L}^{5-(m-1)} = \Lambda_{2}^{5-m} M$.

We now consider the flow $[n,m]\rightarrow [n-1,m]$. Adding a mass term
$M \Lsc{1}{2}$  for one flavor of $\spa$ in the electric theory,
the theory
flows to the
$[n-1,m]$ model, with the scale of $\spa$ given by
$\Lambda_{1 L}^{5-(n-1)} = M\, \Lambda_1^{5-n}$.

In the second dual, upon
adding the term $M \Lsc{1}{2}$  to the
superpotential~(\ref{finalspotential}), the $\Lsc{1}{2}$ equation of motion is:
\beq
\label{l12eom}
p^{\lam_1}\cdot v^1\, J_{\lam_1 \lam_2}\, p^{\lam_2}\cdot v^2\, =\, 2\, M\,
\mu_1\, \mu_2^2\ .
\eeq
The following choice of vevs satisfies~(\ref{l12eom}) as well as the D-flatness
conditions:
\beq
\label{vevs}
p^{\lam = 1}_{\lamd = 1}\, =\, p^{\lam = 2}_{\lamd = 3}\, =\,
v^{i = 1}_{\lamd = 2}\, =\,  v^{i = 2}_{\lamd = 4} \, =\, v \,
\equiv \, \left( -2 M \mu_1 \mu_2^2\right)^{1/4}~,
\eeq
with all other components vanishing,
and breaks the group down to $SP(2(n-1)-4)\times SP(4(n-1)+2m-10)$.
The fields
$p_{\lamd=1,3}^{\lam}$,
$p_{\lamd}^{\lam=1, 2}$ and
$\vv{\lamd}{i=1,2}$
are eaten,
and the
superpotential~(\ref{finalspotential}) gives masses to the broken $\spd$
components
($\lamd = 1\ldots 4$) of $\rr{\lamd}{a}$ and $\vv{\lamd}{\hat i}$, where
${\hat i} = 3\ldots 2n$, and to the broken $\spc$ components ($\lam=1,2$) of
$\G{\lam}{a}$ and
$\Ap{\lam_1}{\lam_2}$.
It also gives rise to masses for the gauge singlets
$\Lsc{i}{j}$, $\Y{i}{a}$, $\W{i}{j}$ with $i$ or $j$ equal to 1  or  2.
After integrating out the heavy fields, the resulting superpotential is of the
form~(\ref{finalspotential}) with $i = 3\ldots 2n$, $\lam = 3\ldots 2n-4$ and
$\lamd = 5\ldots 4n+2m-10$, as expected for the second dual of the $[n-1,m]$
model. The only fields which transform under the unbroken groups and become
massive from the superpotential are the $SP((2 (n - 1) - 4)$ fundamentals
${1 \over \mu_2} A^\prime_{\lambda_1 = 1,2;~  \lambda_2 > 2}$, which  mix
with the $p^{\lambda > 2}_{\lamd = 2,4}$ components. These two flavors of
the unbroken  $SP((2 (n - 1) - 4)$ have equal masses $v/2$.

The scale $\bar{\Lambda}_{1 L}$  of $SP(2 (n - 1) - 4)$  is affected by
two factors under this  flow:
 the $SP(2 n - 4)$ in the high-energy theory is broken to $SP(2 (n - 1) - 4)$
by the expectation values (\ref{vevs}) of the fields $p_{\lamd = 1,3}$,
while at the same time two flavors  --  the fields 
${1 \over \mu_2} A^\prime_{\lambda_1 = 1,2;~  \lambda_2 > 2}$ and 
$p^{\lambda > 2}_{\lamd = 2,4}$ mentioned above  --  of 
 the unbroken $SP(2 (n - 1) - 4)$
gain mass $v/2$ from the superpotential (\ref{finalspotential}).
Using    (\ref{scalematchingflat})  and
(\ref{scalematchingmass}), we can find the matching condition for
$\bar{\Lambda}_1$ for this flow:
\beqa
\label{lambdabar1nflow}
\bar{\Lambda}_{1 L}^{5 - 2m} &=& \left( {v \over 2} \right)^2 ~ {2 \over v^2} ~
 \bar{\Lambda}_{1}^{5 - 2m}
= {1 \over 2} ~c(n,m) ~{ \Lambda_2^{5 - m} \over \mu_1~ \mu_2^{m - 1}
}\nonumber \\
&=& c(n - 1, m) ~{ \Lambda_2^{5 - m} \over \mu_1 ~\mu_2^{m - 1}}~.
\eeqa
Eq. (\ref{lambdabar1nflow}) implies a recursion relation for
$c(n,m)$:
\beq
\label{cmrecursion2}
c(n,m) = 2~ c(n - 1, m)~,
\eeq
which, together with eq. (\ref{cmrecursion}) and (\ref{cof1fixed}) implies that
$c(m,n) = (-1)^m 2^{m + n -3}$.

The scale ${\bar \Lambda}_{2\,  L} $
of the $SP(4(n-1)+2m-10)$ can be found using (\ref{scalematchingflat}):
\beqa
\label{lam2barl}
{\bar \Lambda}_{2\,  L}^{4(n-1)+2m-10}\, &=&
4v^{-4}\,
{\bar \Lambda}_2^{4n+2m-10}
\, =\,
 - { 2 {\bar \Lambda}_2^{4n+2m-10} \over
M\, \mu_1\, \mu_2^2 }
\nonumber \\
 &=& \,
 (-1)^{(n-1)+m} 2^{6-(n-1)} \,
{ \mu_2^{2(n-1)+m-2} \, \mu_1^{2+(n-1)} \over
\Lambda_{1\, L}^{5-(n-1)} \, \Lambda_2^{5-m} } \ ,
\eeqa
which is the correct scale for the second dual of the $[n-1,m]$ model
(equation~(\ref{lambdab2})).

Finally, we consider adding a mass $M $ for the field $X$.
Recall that upon integrating out
the field $X$ in the electric theory, one gets a theory with two separate
$SU(2)$ gauge groups; $\spa$ with $n$ flavors, and $\spb$ with $m$ flavors.

In the second dual, $X$ is not present since it becomes heavy and gets
integrated out. However, the equation of motion~(\ref{eom})
relates it to the light field $p$, which transforms under both $\spc$ and
$\spd$. Thus, the appropriate term to add to the superpotential, eq.
(\ref{finalspotential}), is given by:
\beq
\delta W =\,
{1\over 2n-4}\,
 {\mu_1\over \mu_2}\, M \, J_{\lam_1 \lam_2} p^{\lam_1}\cdot p^{\lam_2}
\nonumber
\eeq
Once it is added the field $p$ becomes heavy and
can be integrated out, so that the $\spc$ and $\spd$ gauge theories are now
decoupled (except for the superpotential). Now however, $\spd$ has $N_f=n+m$
and confines for $n\ge 3${\footnote{For $n\le 2$ the discussion in this section
is not valid. See following sections.}}. One is then left with an $\spc$ gauge
group
with an antisymmetric, $2m$ fundamentals and a number of singlets (including
the mesons of the confining $\spc$). The superpotential of this theory is
partly
given by~(\ref{finalspotential}), after substituting for the field $p$,
from its equations of motion, in terms of
the fields $\Lsc{i}{j}$, $v^i$, $G_{a\lambda}$ and
$\Ap{\lam_1}{\lam_2}$, and going over to the $\spd$ mesons.
In addition, there is a non-perturbative contribution to the superpotential
generated by the confining $\spd$.

The dynamics of this $\spc$ gauge theory is thus very involved. Although we
cannot rigorously argue that this is indeed the case, we 
conjecture that in the infra-red, the 
theory gives two decoupled sectors  --  one 
corresponding to the electric $SU(2)_2$ with its $m$ flavors, and the other
corresponding to an $SP(2n-6)$ gauge theory that is the dual of the electric
$SU(2)_1$.

This picture can be substantiated by considering the special case $n=3$, for
which $\spc=SU(2)$ and $\spd=SP(2m+2)$. Once the field $p$ is integrated out,
$SU(2)$ is left with the $2m$ doublets
$G_{{\lam}a}$ (notice there is no antisymmetric in this case). $SP(2m+2)$, on
the other hand, has $N_f=N_c+2$ and confines,
generating the superpotential  ${\rm Pf} {\tilde M}$ (\ref{spsuperpotential}),
where  ${\tilde M}$
denote collectively the $SP(2m+2)$ mesons $M^{ij}=v^i\cdot v^j$,
$M^{ab}=r^a\cdot r^b$ and $M^{ia}=v^i\cdot r^a$. These mesons become heavy due
to the superpotential~(\ref{finalspotential}); similarly, the fields $Y_{ia}$,
$W_{ij}$ and $\Rsc{a}{b}$ become heavy. Substituting for the fields $p$ as well
as for the other heavy fields, we find a vanishing superpotential. The second
dual therefore gives an $SU(2)$ gauge theory with $m$ flavors and no
superpotential, corresponding to the the electric $SU(2)_2$ (the theories are
not identical, there is no simple field redefinition that relates their
fundamentals). As for  the electric $\spa$, this theory confines for the case
$n=3$ we are considering. The non-perturbative superpotential that this theory
generates probably arises in the dual through some complicated dynamics. Aside
from this superpotential however, the magnetic and electric theory are clearly
identical in the infra-red.

\subsection{Deformations Along Flat Directions.}

\label{deformationsalongflatdirections}
 
As a further check of the duals we constructed, we now analyze their moduli spaces. To do this, various fields in the electric and magnetic theories are given expectation values along $D$- and $F$-flat directions. We then verify that the resulting theories are equivalent in the infra-red. 
In the process we  also show that the 
chiral rings of the electric and magnetic theories  are identical.
 As  is common in the study of duality, classical
restrictions on the chiral ring  of  the electric
theory will
 sometimes arise through complicated, nonperturbative
effects in the dual description.

We limit our discussion to the second dual of 
section~\ref{thedualsofthenmmodels}, since, for the purpose of this analysis, the first dual is essentially the same as simple $SP$ duals.

We  begin with the flat direction parametrized by the expectation 
value of the field $X$. In the electric theory the $X$ expectation value
 breaks the gauge symmetry to the diagonal
$SU(2)_D$. The diagonal gauge group has now $2 n + 2 m$ doublets and a
scale $\Lambda_D^{6 -  n - m} = \Lambda_1^{5 - n} ~ \Lambda_2^{5 - m}/ X^2$.
Its
dual is an $SP(2 n + 2 m - 6)$ gauge theory with $2 n + 2 m$ fundamentals, 
and a
scale obeying:
\beq
\label{scaleindualx}
\bar{\Lambda}_D^{2 n + 2 m - 6}  = { 16 (-)^{n + m } ~\mu_D^{n + m}~ X^2
\over
 \Lambda_1^{5 - n} ~ \Lambda_2^{5 - m} }~.
\eeq
We
will show below that the $SP(2 n - 4) \times SP(4 n + 2 m - 10)$ dual
theory flows to this
dual after deforming it along the $X \ne 0$ flat direction,
thereby establishing its equivalence with the electric theory.

By the equation of motion for the field $X$, eq.  (\ref{eom}), giving an
expectation value to $X$
in the electric theory is equivalent to giving an expectation value to
$p \cdot p$ in the magnetic theory.
The $D$- and $F$-flat
conditions that follow from (\ref{finalspotential}) determine the expectation
values of the field $p$:
\beq
\label{xflat1}
p^{\lam}_{\lamd} = \left\{ \begin{array}{ll}
      \sqrt{ - ~ {\mu_2 ~X \over \mu_1}}~
   \delta^{\lam}_{ \lamd} & \mbox{if $\lamd = 1,...,2 n - 4$}\\
                      0 & \mbox{otherwise}~. \end{array} \right.
\eeq
The $p$ expectation value (\ref{xflat1})
breaks the $SP(2 n - 4) \times SP( 4 n + 2 m - 10)$ gauge
symmetry down to $SP(2 n - 4)_D \times SP(2 n + 2 m - 6)$.
All fundamentals of
$SP(4 n + 2 m - 10)$ decompose into fundamentals of the
unbroken $SP(2 n + 2 m - 6)$, denoted in the following by
hatted fields, and fundamentals of the diagonal unbroken $SP(2 n - 4)_D$,
denoted in the following by barred fields (e.g. the fields $v^i$ decompose
into $\hat{v}^i$ which are  fundamentals
of the unbroken $SP(2 n + 2 m - 6)$, and $\bar{v}^i$ which
are fundamentals of $SP(2 n - 4)_D$).
The field $p$ decomposes under
$SP(2 n - 4)_D \times SP(2 n + 2 m - 6)$ as follows:
$p = ({\bf 1},  {\bf 1}) + (\Ysymm,  {\bf 1}) + (\Yasymm, {\bf 1})  +
(\Yfund, \Yfund)$,
where $\Yasymm$ is traceless.
The  $(\Ysymm,  {\bf 1}) $ and
$(\Yfund, \Yfund)$ are eaten by the Higgs mechanism. The
$(\Yasymm, {\bf 1}) $ part of $p$
 pairs with the field $A^\prime$ and becomes massive
due to the superpotential. Thus only the singlet part of $p$ remains
massless, and corresponds to
the field $X$.
 Furthermore, by inspecting the superpotential (\ref{finalspotential}) after
substituting
the vevs (\ref{xflat1}), we observe that the fields $G_a$ and ${\bar r^a}$
become massive and
can be integrated out. After integrating out the heavy fields the resulting
superpotential is:
\beqa
\label{xflat2}
W &=& {1 \over 4~ \mu_1^2 ~\mu_2}~ (  W_{ij} - X ~ \scL_{ij} )~
\bar{v}^i \cdot \bar{v}^j +
{1 \over 4~ \mu_2}~\scR_{ab}~
\hat{r}^a \cdot \hat{r}^b \nonumber \\
 &+& {1 \over 2~ \mu_1 ~ \mu_2 } ~ Y_{ia} ~ \hat{v}^i \cdot \hat{r}^a  +
 {1 \over  4 ~ \mu_1^2 ~\mu_2} ~ W_{ij} ~ \hat{v}^i \cdot \hat{v}^j  ~.
\eeqa
The diagonal $SP(2 n - 4)_D$ now has  only $2 n$ fundamentals, the
fields $\bar{v}^i$, and is
therefore confining, generating a nonperturbative superpotential
(\ref{spsuperpotential})
$W_{n.p.} \sim {\rm Pf} M$,  with
$M^{ij} \equiv \bar{v}^i \cdot \bar{v}^j $ being the confined degrees of
freedom.
 Adding $W_{n.p.}$ to the superpotential (\ref{xflat2}) we obtain:
\beqa
\label{xflat3}
W &=& {1 \over 4~\mu_1^2 ~\mu_2}~ (  W_{ij}  - X ~ \scL_{ij}  )~ M^{ij}+
{1 \over 4~\mu_2}~\scR_{ab}~
\hat{r}^a \cdot \hat{r}^b \nonumber \\
 &+& {1 \over 2~\mu_1 ~ \mu_2}~ Y_{ia} ~\hat{v}^i \cdot \hat{r}^a  +
 {1 \over 4~\mu_1^2 ~\mu_2}~ W_{ij} ~\hat{v}^i \cdot \hat{v}^j  -
{ {\rm Pf} M \over 2^{n - 3} ~\bar{\Lambda}_{D}^{ 2 n - 3} }~,
\eeqa
where $\bar{\Lambda}_{D}^{ 2 n - 3}$ is the scale of the diagonal $SP(2 n -
4)_D$.
One linear combination of $W$ and $\scL$ obtains mass together with the meson
$M$ and can be integrated out.
The equations of motion for the heavy fields imply 
 $M^{ij} = 0$ and $W_{ij} = X \scL_{ij}$. Note that the second equality
reproduces a
classical constraint in the electric theory.
Integrating the heavy fields out, the
superpotential becomes:
\beqa
\label{xflat4}
W =
{1 \over 4~\mu_2}~\scR_{ab}~
\hat{r}^a \cdot \hat{r}^b
 + {1 \over 2~\mu_1 ~ \mu_2}~ Y_{ia} ~\hat{v}^i \cdot \hat{r}^a  +
 {1 \over 4~\mu_1^2 ~\mu_2}~ X~\scL_{ij}~\hat{v}^i \cdot \hat{v}^j ~.
\eeqa
Eq.(\ref{xflat4}) can be identified with the superpotential of the
$SP(2n+2m-6)$ dual of the electric $SU(2)_D$ theory mentioned above.
To see this note that once $X$ acquires a vev the fields transforming
under $SU(2)_D$  can be taken to be $R_a$ and ${Q\cdot L_i \over \sqrt{X}}$.
On dualizing
$SU(2)_D$ with this matter content one gets a
superpotential given precisely by eq.(\ref{xflat4}),
with $\mu_1 $
set equal\footnote{
Equivalently
we can perform the $X$- and $\mu_1$-dependent field redefinition $Y \rightarrow
Y \sqrt{X}$,
 $\hat{v} \rightarrow \hat{v} \mu_1/\sqrt{X}$, so that the fields have the
same symmetries
as those in the dual of  $SU(2)_D$. As a result $\mu_1$ disappears from the
superpotential
and the scale matching relation.} to $\sqrt{X}$. Thus, the electric
and the dual
theory match along the $X \ne 0$ flat direction.

Next we consider the flat direction $\scR_{ab} \ne 0$.
We begin by giving $\scR_{ab}$ a rank $2$ expectation value.
For the sake of definiteness we take $\scR_{12} \ne 0$. This expectation
value completely higgses $SU(2)_2$.
  The low-energy  electric theory is
then the $SU(2)_1$ gauge theory with $2 n + 2$ doublets ($Q_{\ald}, L_i$),
and the massless singlets
$\scR_{a \hat{a}}$  --  the moduli describing the
flat direction and the components of the $SU(2)_2$ matter fields that are not
eaten.  In the following,
the indices $\hat{a}, \hat{b} = 1, 2$. By the usual rules
of $SP$ duality,  this low-energy
$SU(2)_1$  has a dual
description in terms of an $SP(2n-4)$ theory. We will see below that
the magnetic theory correctly reduces to this $SP(2n-4)$ theory along
the $R_{12} \ne 0$ direction.

In the dual theory, the expectation value of $\scR_{12}$ gives
mass to $2$ fundamentals of $SP(4 n + 2 m  - 10)$.  The low-energy
$SP(4 n + 2 m  - 10)$ theory (whose scale will be denoted by
$\bar{\Lambda}_{2_L}$ ) now  has $4 n + 2 m - 6$ fundamentals.
It is therefore confining,  and generates
a nonperturbative superpotential (\ref{spsuperpotential}), $W_{n.p.} \sim {\rm
Pf} {\cal M}$.
Here ${\cal M}$ denote
the mesons of the confining $SP(4 n + 2 m  - 10)$:
 $M^{i \lam} = v^i \cdot p^{\lam}, ~M^{i j} = v^i \cdot v^j,
 ~M^{\lambda \lambda^\prime} =
p^{\lam} \cdot p^{\lambda^\prime}$,
 $M^{a^\prime b^\prime} = r^{a^\prime} \cdot r^{b^\prime}$ (with $a^\prime,
b^\prime = 3,..., 2 m$ only), $M^{a^\prime i} = r^{a^\prime} \cdot v^i$ and
$M^{a^\prime \lambda} = r^{a^\prime} \cdot p^{\lambda}$.
We now integrate out the heavy flavor $r^{\hat{a}}$ from
(\ref{finalspotential}),
add the nonperturbatively generated
superpotential along the $\scR_{12} \ne 0$ flat direction, and
rewrite the resulting superpotential in terms of the mesons defined above,
much like we did when considering the $X \ne 0 $ flat direction:
\beqa
\label{rflat1}
W &=& - {{\rm Pf} {\cal M} \over 2^{2 n + m - 6} ~
\bar{\Lambda}_{2_L}^{4 n + 2 m - 9} }
 - { 1 \over 4 \mu_1 ~\mu_2^2}~ {\scL}_{ij} ~
 M^{i \lambda} ~ M^j_{\lambda}  \nonumber \\
&+&
{ 1 \over 4 ~\mu_2 ~{\scR}_{12} }~
\left(
- {\cal R}_{\hat{a} a^\prime} ~ M^{a^\prime b^\prime} ~ {\cal
R}^{\hat{a}}_{b^\prime}
+ 2 ~ {\cal R}_{\hat{a} a^\prime} ~ M^{a^\prime \lambda} ~
G^{\hat{a}}_{\lambda}
+ {2 \over \mu_1} ~ {\cal R}_{\hat{a} a^\prime} ~ M^{a^\prime i} ~
Y_i^{\hat{a}}
\right. \nonumber \\
& ~ & \left. \hspace{2.5cm}
 - G_{\lambda \hat{a}} ~ M^{\lambda \nu} ~ G_{\nu}^{\hat{a}}
- {2 \over \mu_1} ~ G_{\lambda \hat{a}} ~ M^{\lambda i} ~ Y_i^{\hat{a}}
- {1 \over \mu_1^2}~Y_{i \hat{a}} ~ M^{ij} ~ Y_j^{\hat{a}}
 \right) \\
&+& {1 \over 4~ \mu_2}~
\left(
2~ G_{\lambda a^\prime} ~ M^{\lambda  a^\prime}
+ {2 \over \mu_1} ~ Y_{i a^\prime}~M^{i a^\prime}
+ {\cal R}_{a^\prime b^\prime} ~ M^{a^\prime b^\prime}
+  A^\prime_{\lambda_1 \lambda_2} ~ M^{\lambda_1 \lambda_2}
+ {1 \over \mu_1^2} ~ W_{ij} ~ M^{ij}
\right) ~.\nonumber
\eeqa
{}From the superpotential
eq.(\ref{rflat1}) we see that the only fields that remain massless are
the $SP( 2 n - 4)$ fundamentals $G_{\lambda \hat{a}}, M^i_{\lambda}$ and the
gauge
singlets
$\scL_{ij}, Y_{\hat{a} i}, \hat{M} = M^{\lambda \lambda^\prime}
J_{\lambda \lambda^\prime}$ and
  $\scR_{a \hat{a}}$. The nonperturbative
superpotential $W_{n.p}$ vanishes after imposing  the equations of motion for
the heavy fields, and
the superpotential of the  remaining massless fields is:
\beq
\label{wrflat}
W  =   - {1 \over 4 ~ \mu_1 ~ \scR_{12} } ~ X ~ \varepsilon_{\hat{a} \hat{b}}
{}~  G_{\lambda}^{\hat{a}} ~ G^{\lambda \hat{b}}
 + {1 \over 2 ~ \mu_1 ~ \mu_2  ~ \scR_{12} } ~ Y_{\hat{a} i} ~
 G^{\hat{a}}_{\lam} ~ M^{i \lam} +
{1 \over 4 ~ \mu_1~ \mu_2^2} ~ \scL_{ij} ~ M^{i}_{\lambda} ~ M^{j \lambda}~,
\eeq
where we used the relation $\hat{M} = (2 n - 4)  \mu_2 X/ \mu_1$ (\ref{eom}).

But this superpotential, as promised,  is precisely the superpotential of the
dual of the low-energy
electric $SU(2)_1$ with $2 n + 2$  doublets, eq. (\ref{dual1spotential}).
To see this note that once $\scR_{12} \ne 0$, the matter fields transforming
under $SU(2)_1$ in the electric theory can be taken to be $L_i$ and
${R_a \cdot Q \over \sqrt{\scR_{12} } }$. Eq. (\ref{wrflat}) suggests
that the field dual to  $L_i$ be identified with
$l^i_\lambda \equiv {1 \over \mu_2} M_{\lambda}^i$. In addition, on
identifying the field dual to
${R_a \cdot Q \over \sqrt{\scR_{12} } }$ with
$q^{\hat{a}}_{\lam} \equiv
{G^{\hat{a}}_{\lam} \over \sqrt{\scR_{12}} }$
and $V_{\hat{a} i} \equiv {Y_{\hat{a} i} \over \sqrt{\scR_{12}}}$
we find that eq. (\ref{wrflat})
agrees with the required superpotential of the dual of the low-energy
$SU(2)_1$.
This agreement shows that the electric and dual theory agree along
$\scR_{12}$ flat direction too.


Classically, the chiral ring of the electric theory satisfies 
 a number of relations. These ensure, for example,  
that $\scR_{ab}$ cannot have an expectation value of rank greater than 2.
In order to see how these relations arise in the dual theory let us return to
eq. (\ref{rflat1}).
As was mentioned above, several fields get mass due to bilinear
couplings in this superpotential. Integrating them out gives rise
to equations relating the heavy fields with the light ones and these,
in fact,  correctly
reproduce the relations in the chiral ring of the electric theory.
For example,  the field $ M^{a^\prime b^\prime} $ gets a mass by pairing
with ${\cal R}_{a^\prime b^\prime}$. On integrating it out we find that
\beq
\label{cconstr}
{\cal R}_{a^\prime b^\prime} = {1 \over {\cal R}_{12}}~
        {\cal R}_{\hat{a} a^\prime} ~{\cal R}^{\hat{a}}_{b^\prime}\ .
\eeq
These are the relations in the chiral ring of the electric theory
which result in $\scR_{ab}$ having an expectation value of rank $\le 2$.
Notice that, whereas  in the electric theory  these are classical relations,
in the dual  they  arise after including the
effects of confinement, as  in (\ref{rflat1}).

We can continue turning on
further expectation values in eq.(\ref{wrflat}). For example,
an $\scL_{12}$ expectation value
gives mass to two $SP(2 n - 4)$ fundamentals, so that this group confines
(recall that along this
flat direction the electric theory is completely higgsed).
It is easy to work out the massless
spectrum that follows from eq.~(\ref{wrflat}),
after accounting for the nonperturbative
superpotential generated by the confining $SP(2 n - 4)$,
and to see that it precisely matches
that of the electric theory. The superpotential in terms of the
massless fields vanishes in both theories too.
Thus, along the $\scL_{12},\scR_{12} \ne 0$
 flat direction the electric and magnetic theories   flow
to the same (trivial) infra-red theory. Furthermore, when the analysis of
the chiral ring is extended to this case, one finds that the magnetic
theory reproduces all the constraints in the electric theory,
so that the  chiral rings in the two cases are the same.

This concludes our discussion of the flat directions in the
electric and dual theories. As we have seen, the behavior of
these theories agrees along various flat directions. This provides
strong additional evidence for their equivalence at low energies.

\subsection{Flows by Yukawa Perturbations.}

\label{flowsbyyukawaperturbations}

In this section we discuss briefly the flows due to Yukawa
perturbations in the
second dual. We add
the term
\beq
\label{wyukawa}
W = \lambda^{ia}~ Y_{ia}
\eeq
to the superpotential (\ref{finalspotential}) with a Yukawa-coupling
matrix of rank $ P \le {\rm min} \{ 2m, 2n \}$. Upon adding the perturbation
(\ref{wyukawa}), the
dual $SP(2n-4)\times SP(4 n + 2 m -10 )$
 theory flows to a new fixed point.
The fields  $v^i$ and $r^a$  get
 expectation values that obey the
F-flatness conditions
\beq
\label{flatconditionsYuk}
v^i \cdot r^a  = - 2 ~\mu_1 ~\mu_2~ \lambda^{ia} , ~ ~ ~ r^a \cdot r^b = v^i
\cdot v^j = 0~,
\eeq
and,  along the D-flat directions, can be taken to be:
\beq
\label{yukawavevsv}
v^i_{\lamd} = \left( \begin{array}{ccccccc}
\sqrt{s_1}&0& \ldots & 0 & 0 & \ldots &0 \\
0&0&\ldots&0& 0 & \ldots &0 \\
0&\sqrt{s_2}&\ldots&0& 0 & \ldots &0 \\
0&0&\ldots&0& 0 & \ldots &0 \\
\ldots&\ldots&\ldots&\ldots&\ldots&\ldots&\ldots \\
0&0&\ldots&\sqrt{s_P}&0& \ldots &0 \\
0&0&\ldots&0& 0 & \ldots &0\\
\ldots&\ldots&\ldots&\ldots&\ldots&\ldots&\ldots \\
0&0&\ldots&0& 0 & \ldots &0 \end{array} \right)~,
\eeq
\beq
\label{yukawavevsr}
r^a_{\lamd} =\left( \begin{array}{ccccccc}
0&0& \ldots & 0 & 0 & \ldots &0 \\
\sqrt{s_1}&0&\ldots&0& 0 & \ldots &0 \\
0&0&\ldots&0& 0 & \ldots &0 \\
0&\sqrt{s_2}&\ldots&0& 0 & \ldots &0 \\
\ldots&\ldots&\ldots&\ldots&\ldots&\ldots&\ldots \\
0&0&\ldots&0&0& \ldots &0 \\
0&0&\ldots&\sqrt{s_P}& 0 & \ldots &0\\
\ldots&\ldots&\ldots&\ldots&\ldots&\ldots&\ldots \\
0&0&\ldots&0& 0 & \ldots &0 \end{array} \right)~,
\eeq
where by a field redefinition we took the rank-$P$ Yukawa matrix to be
diagonal, with
eigenvalues $\lambda_1 \ldots \lambda_P$ and $s_i = - 2 \mu_1 \mu_2 \lambda_i$;
in eqs. (\ref{yukawavevsv}), (\ref{yukawavevsr}) the $SP(4 n + 2 m - 10)$
gauge indices $\lamd$ are taken to enumerate the rows.

The expectation values (\ref{yukawavevsv}), (\ref{yukawavevsr}), 
higgs the dual theory to  $SP(2 n-4) \times SP(4 n + 2 m -10 - 2 P) $.
We will not consider  in general the new fixed point, but will only note the
interesting case when the Yukawa  coupling has rank $P = 2 n + m - 5$.
In this case,
the $ SP(4 n + 2 m -10) $ group is completely broken, while the matter
content of the
$SP(2 n - 4)$ theory consists of the antisymmetric tensor $A^\prime$,
the $2 m - P$ fundamentals $G_{a \lambda}$ with $a > P$,
and $2 n + m - 5$ of the components of $p_{\dot{\lambda}}$
(half of the components
of $p_{\dot{\lambda}}$ and the
components of $G_{a \lambda}$ with $a \le P$
become heavy, as can be seen by substituting the expectation
 values of
$v^i$ and $r^a$ in the superpotential (\ref{finalspotential})).
Symmetry considerations show that this theory confines
for  $m \le 3$; the $m=3$ case exhibits confinement without
chiral symmetry breaking.

We will not analyze the confining phase of this theory in
detail in this paper,
our only purpose here is to  note
that it is  possible 
to flow to the confining phase
by perturbing the superpotential  with dimension-3 terms,
without having to add mass terms for any field.
In the more interesting chiral product-group theories,
such flows to the
confining phase might be interesting from the
point of view of supersymmetry breaking. We leave the detailed
investigation of this
for future work.

\mysection{The ``Partially Confining" Models.}

\label{thepartiallyconfiningmodels}

\subsection{The ${\bf [2,m]}$ Models.}

\label{the2mmodels}

In this section we study  theories in which one of the
two electric gauge groups
is in the confining regime. By this we mean, more precisely, that one of the
two
groups, say $SU(2)_1$, has three or fewer flavors and would therefore confine
in the absence of $SU(2)_2$. These theories have a rich set of non-perturbative
phenomena which duality helps elucidate.
This section deals with the $[2,m]$ models, and the following two
sections deal with the $[1,m]$ theories.

A convenient starting point for studying   the $[2,m]$ models is
 the limit $\Lambda_1 \gg \Lambda_2$,
in which $SU(2)_1$ confines at the scale $\Lambda_1$.
Below this scale one can use an effective theory in terms of the $SU(2)_1$
mesons,
$X\equiv Q^2$,
$\Lsc{i}{j} \equiv  L_i \cdot L_j$  and
$\V{\ald}i \equiv Q_{\ald} \cdot L_i$ with $i,j=1\ldots 4$.
 Non-perturbative effects in the confining group
give rise to a  superpotential (\ref{spsuperpotential})
\beq
\label{npsp}
W=-{1\over \Lambda_1^3 }~ \left( X ~{\rm Pf} {\cal L} - {1\over 4}~ W_{ij}  \
\Lsc{k}{l} \
                         \varepsilon^{i j k l } \right) \ ,
\eeq
with $W_{ij}=V_i \cdot V_j $ as in section~2.
$SU(2)_2$ in this effective theory has $2m+4$ doublets,
$\R{\ald}a$ and $\V{\ald}i/\Lambda_1$,
 and its scale is given by\footnote{A constant could in principle appear in
this scale matching
relation, multiplying the right hand side. However it can be shown to be $1$,
by first establishing, by considering the $m \rightarrow m-1$ flow,
that it is $m$-independent, and then evaluating it in the confining $[2,1]$
model by adding mass terms, calculating the vevs of
various fields and  demanding consistency with the Konishi anomaly.}:
\beq
\label{sce}
\Lambda_{2^\prime}^{4-m} = {\Lambda_2^{5-m} \over \Lambda_1}\ .
\eeq

The dual theories are best understood by starting with the $[3,m]$ duals and
flowing to the $[2,m]$ case after adding a mass term for one
$L$ flavor. In the previous section we presented two types of $[3,m]$ duals:
one where only one $SU(2)$ is dualized, and  the other where both $SU(2)$s are
dualized.
In the former,
the $[3,m] \rightarrow [2,m]$ flow is very similar to the
 $N_f=3 \rightarrow N_f=2$ flow
in  basic $SP$ duality~\cite{sp}. We therefore only discuss here duals of the
second type, in which both groups are dualized.
There are two such theories. Dualizing $SU(2)_1$ followed
by $SU(2)_2$, a $[3,m]$  dual with gauge group
 $SU(2)\times SP(2m+2)$  is obtained. Reversing this order, an
$SP(2m-4) \times SP(4m-4)$ theory is obtained instead.

We first discuss the $[3,m]\rightarrow [2,m]$ flow in the
 $SU(2)\times SP(2m+2)$ dual.
  The analysis of this flow has much in common with that of
section 2.2.2. Therefore we only point out the
essential differences here.

 Adding a mass term  $M \Lsc{1}{2}$  to the
superpotential~(\ref{finalspotential}) of
the $[3,m]$ theory, the $SU(2) \times SP(2m+2)$ dual gauge group gets broken to
its $SP(2m-2)$ subgroup. Several fields get heavy either through the Higgs
mechanism or by pairing with other fields through couplings in the
superpotential~(\ref{finalspotential}).
The fields that remain light  are
$\rr{\lamd}a$, $\vv{\lamd}i$, which transform under the
$SP(2m-2)$ gauge symmetry, and $\Lsc{i}{j}$, $\Rsc{a}{b}$, $\Y{i}{a}$,
$\W{i}{j}$ and $X$ ($i,j = 3...6$), which are gauge singlets\footnote{
The field $X$ arises as follows: unlike the general $n$ case, the $n=3$
$SU(2) \times SP(2m+2)$ theory does not have the
$\Ap{\lam_1}{\lam_2}$ field. As a result, one
component of $\p\lam\lamd$ does not get mass and can be identified
with $X$.}.
The resulting superpotential is:
\beq
\label{superpottwo}
W = {1\over 4 {\mu_2}}~\left( \Rsc{a}{b}~ r^a \cdot r^b +
{1\over \mu_1^2 } ~ \W{i}{j} ~ v^i\cdot v^j
+ {2 \over \mu_1} ~ \Y{i}{a} ~ v^i \cdot r^a \right) \  .
\eeq

Since the $SU(2)$ subgroup is completely broken, one might
expect the non-perturbatively
generated superpotential~(\ref{npsp}) to arise
from an instanton in this subgroup. However,  careful consideration
of the zero-modes involved shows that the contribution from this instanton
vanishes. Furthermore, the scale matching relations eqs.~(\ref{lambdab1}) and
(\ref{lambdab2}), with $n = 3$,
imply the following relation between the scale $\Lambda_1$  of the
 $[2,m]$ electric theory
($\Lambda_1^3 = M \Lambda_{1 H}^2$, with $\Lambda_{1 H}$ the corresponding
scale in the $[3,m]$ theory)
and the scales  $\bar{\Lambda}_1$ and $\bar{\Lambda}_2$
of the $[3,m]$ magnetic theory:
\beq
\label{smntwo}
{1 \over \Lambda_1^3} \sim
{ \bar{\Lambda}_1^{5-2m} ~ \bar{\Lambda}_2^{2m+2}
  \over   M ~ \mu_1^4 ~ \mu_2^5 } ~ .
\eeq
Hence, since $\Lambda_1$ is the scale appearing in (\ref{npsp}),
this nonperturbative term
 must arise from instanton-like
configurations with one unit of winding in both the $SU(2)$ and the $SP(2m+2)$
subgroups. While we have not actually calculated
this contribution,
 the counting of zero modes
suggests that it can arise in this manner. We expect these
configurations to include, but not be restricted to, instantons that
lie in a diagonal $SU(2)$ subgroup of $SU(2) \times SP(2m+2)$.

This non-perturbative effect is different from those encountered in
 the simple-group case, in that it arises from configurations with
components lying in {\it partially} broken subgroups. It  should be a generic
feature of product group theories.

Once we accept that (\ref{npsp}) does arise, we find that the resulting theory
is the expected dual of the electric theory
(in the limit $\Lambda_1 \gg \Lambda_2$ ) with which we started. Below the
scale
$\Lambda_1$, the electric theory was
$SU(2)_2$ with $2m+4$ doublets,  whose dual should be an
$SP(2m-2)$ theory. This is precisely the theory we find by flowing down from
the $[3,m]$ dual.
 The required matter fields and
superpotential are also in agreement with
those found above in the dual theory, provided we take $\mu_1=\Lambda_1$.
Note that the scale
$\mu_1$  in eq. (\ref{superpottwo}) arises because $ Y_{i a}/\mu_1$
and $W_{i j}/ \mu_1^2$ are the canonically normalized fields. In  the
electric theory,
${V_{{\ald}i}/ \Lambda_1}$ are the correctly normalized $SU(2)_2$
doublets,
consequently, $\mu_1$ is set equal to $\Lambda_1$.
In fact, in the
$SP(2)\times SP(2m+2)$ dual discussed above, the strong coupling scale of
$SP(2m+2)$ is given
by~(\ref{lambdab2}), with $n=3$ and $\Lambda_1 \rightarrow \Lambda_{1 H}$.
Flowing down to the $[2,m]$ dual, we find that the $SP(2m-2)$ scale is given by
\beq
\label{scmta}
\bar{\Lambda}_2^{2m-2}=16  \ (-1)^m \
                        { \mu_2^{ 2 + m } ~\mu_1^4 \over \Lambda_2^{5-m}
                           ~ \Lambda_1^3 }\ ,
\eeq
or, setting  $\mu_1=\Lambda_1$,
\beq
\label{scmtb}
\bar{\Lambda}_2^{2m-2}=16 \ (-1)^m \
                         { \mu_2^{2+m} ~ \Lambda_1 \over \Lambda_2^{5-m}}\ ,
\eeq
which is in agreement with~(\ref{sce}) and   the $SP$-duality
matching relation (\ref{spscalematching}):
\beq
\Lambda_{2^\prime}^{4-m}~ \bar{\Lambda}_2^{2m-2} = 16  \ (-1)^m \
                         \mu_2^{2+m} \ .
\eeq

We end our description of this dual theory with one final comment.
 From the  point of view of the electric theory,
we could, strictly speaking, justify the above-mentioned
description of low-energy physics only in the limit
 $\Lambda_1\gg\Lambda_2$. The dual description on the other hand is
valid for all values of the ratio $\Lambda_1/\Lambda_2$.
The equivalence between the dual and electric theories implies
then that the electric description too must be more generally valid.

As mentioned above, the $[3,m]$ theory also has a dual with gauge group
$SP(2m-4) \times SP(4m-4)$. In this case, the flow to the $[2,m]$ theory
follows the
discussion in Section  \ref{massflowsintheseconddual}
very closely. The resulting theory has an $SP(2m-4)
\times SP(4m-6)$
symmetry.
In particular, none of the
groups is completely broken, and no non-perturbatively
generated superpotential is expected to arise.
In spite of this, by studying its flat directions as in
Section \ref{deformationsalongflatdirections}
and by deforming the theory after adding mass terms,
as in Section \ref{massflowsandscalematching} one can
show that the
infra-red behavior of this theory
is identical to that of the
electric theory.

\subsection{Dynamically-Generated Dilaton: ${\bf [2,m] }$ 
${\bf \rightarrow [1,m]}$.}

\label{theflow2m1m}

We continue our discussion of the partially confining models by
turning next to the $[1,m]$ case. We will see that
in these theories the electric theory itself
has some interesting features. Specifically, we find that a dilaton is
dynamically generated
in the low energy electric theory and we will see how it can be understood,
in more conventional terms, by flowing from a $[2,m]$ dual theory to a
$[1,m]$ dual theory.  In the subsequent section we will
then study in detail the equivalence between the electric theory and its
various duals.

Let us start, as in the previous section, by considering the electric theory
in the limit $\Lambda_1 \gg\Lambda_2$. At an energy scale of order
$\Lambda_1$ one can go over to an effective theory in terms of the
mesons of $SU(2)_1$,
$X$, $\V{\ald}i$, and $\Lsc{i}{j}$, with a
non-perturbative superpotential
given by:
\beq
\label{qconstr}
W= A~ ( ~X ~\Lsc{1}{2} - W_{12} - \Lambda_1^4~).
\eeq
We can now consider the
effects of the $SU(2)_2$ group in this effective theory. In particular,
one would like to know how its strong coupling scale
is related to that in the microscopic theory.
At first sight it might seem that these two
are equal since the group has $2 m + 2$ fundamentals both in the
ultraviolet theory and
 in the effective theory,  where $\V{\ald}i$ contributes
two fundamentals. However, a little thought involving the symmetries in the
problem shows that this cannot be the case. In fact the effective
theory presents us with a puzzle. There is a  non-anomalous
symmetry in the high energy theory under which $Q$ has charge $1$,
each $L$ has charge $-1$ and each $R$ has charge $-{1\over m}$.
But in the effective theory this symmetry is anomalous, since the
field $\V{\ald}i$  has charge $0$ under this symmetry. How can this
be possible? The answer lies in the
Green-Schwarz anomaly cancellation mechanism. Let us consider those
points in moduli space where $\V{\ald}i$ is zero and the $SU(2)_2$
symmetry is unbroken. We  see from
eq. (\ref{qconstr}) that at such points the
quantum modification of the constraint will force $X$ and $\Lsc{1}{2}$
to acquire vevs  which break the global symmetry described above.
The corresponding Goldstone boson then enters the low energy theory
as an axion and an appropriate shift in this field, along with the
rotations of the $R$ fields, is then a non-anomalous symmetry of the
theory. Since this is a supersymmetric theory, the partner of the axion field
acts as a dilaton. Because of this axion-dilaton field,
the strong coupling scale of $SU(2)_2$ at low-energy
is related to
its value in the microscopic theory in a field-dependent way.
Note that the quantum deformation of the superpotential played
a crucial role in the discussion above. As a consequence, from the point of
view of the electric theory, we can regard the origin of the dilaton as
a truly dynamical effect.

Now let us be more specific. Symmetry considerations tell us that
 the strong coupling scale in the low-energy theory
is given by:
\beq
\label{scmatch}
 \Lambda_{2L}^{5-m} = \Lambda_2^{5-m} ~{\Lsc{1}{2} \over \Lambda_1^2} ~
                    f( {X \Lsc{1}{2}\over \Lambda_1^4} ) ~,
\eeq
where $f({X \Lsc{1}{2}\over \Lambda_1^4}) $ above is an arbitrary function.
We see below how duality
will help determine it completely\footnote{This dependence can be fixed
in other ways, too.  For example
one can flow down to  the [1,2] model, give masses to the different fields
and ensure that the vevs are in  accord with the Konishi
anomaly.}.
In the process we will also  find that  the
dual theory provides a much more
straightforward explanation for the field dependence of the coupling:
it arises because, as usual, the dual of the $[1,m]$ theory is obtained 
by higgsing
the dual of the $[2,m]$ theory. However, in this case, the scale of the
resulting $[1,m]$ theory is not uniquely determined and  depends on a
modulus. This is the required dilaton in the electric theory.

Let us discuss this in  more detail now.
We start with the $[2,m]$ model. The dual which is useful to consider
has a gauge group $SP(2m-2)$ and was discussed at some length
in the previous section. Here we
take $\mu_1$ in eq.(\ref{superpottwo}) to be $\Lambda_1$ and normalize
the $\W{i}{j}$ and $\Y{i}{a}$ fields accordingly.
The superpotential in the
dual theory is given by the sum of (\ref{npsp}) and (\ref{superpottwo}) and is:
\beq
\label{sumsptwo}
W={1\over 4{\mu_2}} ~\left( \Rsc{a}{b}~ r^a \cdot r^b + {1 \over \Lambda_1^2}
                      \W{i}{j}~ v^i \cdot v^j + {2 \over \Lambda_1} ~\Y{i}{a}~
                    v^i \cdot r^a \right)
  - {1\over{\Lambda_1^3}}~ \left( X ~ {\rm Pf}  {\cal L} -  {1\over 4}~ W_{ij}
{}~    \Lsc{k}{l}~ \varepsilon^{i j k l } \right)~.
\eeq
Now we add a mass term for one flavor of the $L$ field:
\beq
\label{delm}
\delta W = m^{3 4}~ \Lsc{3}{4} ~.
\eeq
The equation of motion for $\Lsc{3}{4}$ then reproduces the expected
quantum modified constraint:
\beq
\label{qconstr2}
X\Lsc{1}{2}- W_{12} -\Lambda_1^3 m^{3 4} = 0~,
\eeq
while the equation of motion for $\W{3}{4}$  is:
\beq
\label{veveq}
{\Lsc{1}{2} \over \Lambda_1} + {1\over 2\mu_2} ~v^1 \cdot v^2 =0 ~.
\eeq
The last equation implies that the $SP(2m-2)$ theory is broken to a
$SP(2m-4)$ subgroup. We see that  the scale
${\bar \Lambda_{2L}}$
 of the
low-energy $SP(2m-4)$ theory, which is  given by (\ref{scalematchingflat}),
depends on a modulus,  $\Lsc{1}{2}$, and is:
\beq
\label{sm1}
{\bar \Lambda_{2L}}^{2m-4}={\bar \Lambda_2^{2m-2}}
                            ~\left( -{\Lambda_1 \over \mu_2 \Lsc{1}{2}}
\right).
\eeq
Substituting for  ${\bar \Lambda_2}$ from eq. (\ref{scmtb}) we find
that
\beq
\label{scmtc}
{\bar \Lambda_{2L}}^{2m-4}=16 (-1)^{m+1} \mu_2^{1+m} {\Lambda_1^2 \over
                             \Lambda_2^{5-m} \Lsc{1}{2}} ~.
\eeq
This is consistent with the standard scale matching relation
(\ref{spscalematching}),
applied to $\Lambda_{2L}$ and ${\bar \Lambda_{2L}}$
 only if the scale $\Lambda_{2L}$ of the low-energy
$SU(2)_2$ in the electric theory is given
by:
\beq
\label{fsm1}
 \Lambda_{2L}^{5-m}=\Lambda_2^{5-m} ~{\Lsc{1}{2} \over \Lambda_1^2}~.
\eeq
On comparing with eq. (\ref{scmatch}) we see that this determines the function
$f$ to be a constant equal to $1$.

We end this section with one final comment. As in the $[2,m]$ models,
the low-energy description of the electric theory used above could be justified
only in the limit when $\Lambda_1 \gg \Lambda_2$.
However, the $SP(2m-4)$ theory obtained above is valid for all
values of the ratio $\Lambda_1 /\Lambda_2$. Duality therefore
allows us to conclude that this description of the electric
theory must be more generally valid.

\subsection{The ${\bf [1,m]}$ Models.}

\label{the1mmodels}

We will continue our study of the $[1,m]$ models in this section by
establishing in some detail the equivalence of the electric and
dual theories in the infra-red.

Let us begin by summarizing the important features of the
electric theory. As we saw in the previous section
the low-energy properties of the
electric theory can be described in an effective theory
 consisting of the mesons\footnote{Hereafter, by  $\scL$ we mean $\scL_{12}$},
$X$, $\scL$, $V_{1 \ald}$ and $V_{2}^{\ald}$, with a superpotential eq.
(\ref{qconstr}):
\beq
\label{w1n1}
W = A~ ( X {\cal L} - V_{1 \ald} V_{2}^{\ald}
- \Lambda_1^4 )~.
\eeq
The  scale  of $SU(2)_2$ in this effective
theory,  $\Lambda_{2 L}$, is given by eq. (\ref{fsm1})
(i.e. eq. (\ref{scmatch}) with $f \equiv 1$).

For $m > 2$,  $SU(2)_2$ is in the nonabelian Coulomb phase and
has a dual description in terms of an $SP(2 m - 4)$ gauge group.
This was the dual theory considered in the previous section. The matter
fields in this theory are
$2 m + 2$ dual quarks $v^i, (i = 1,2); r^a, (a = 1,...,2 m)$,
and the gauge singlet mesons ${\cal R}_{ab}$,
 ${1 \over \Lambda_1} Y_{i a}$ and ${1 \over \Lambda_1^2} W_{12}$, with
$W_{12}$ corresponding to the electric meson $V_{1 \lamd} V_{2}^{\lamd}$. The
superpotential of the dual theory, after solving the constraint
(\ref{w1n1}) for the
gauge singlet meson $W_{12}$,  becomes \cite{sp}:
\beq
\label{w1n2}
W = {1 \over 4 \mu_2} ~{\cal R}_{ab}~ r^a \cdot r^b +
 {1 \over 2 \mu_2 \Lambda_1}~ Y_{i a}~ v^i \cdot r^a +
 {X {\cal L} - \Lambda_1^4  \over 4 \mu_2 \Lambda_1^2}~
 v_i  \cdot v^i.
\eeq
The scale of the dual theory $\bar{\Lambda}_2$ is determined by the matching relation
 (\ref{spscalematching})
for $SP$-duality:
\beq
\label{n1scalematching}
\bar{\Lambda}_2^{2 m - 4} \Lambda_{2 L}^{5 - m} =
16 ~(-)^{m + 1} \mu_2^{m + 1} ~.
\eeq

We will consider the moduli space of this dual theory in some detail.

 Consider first those points where $X {\cal L} \ne \Lambda_1^4$. At such
points, it follows from
 eq. (\ref{w1n2}) that
two of the dual quarks, $v^i$, are massive, and
$SP(2 m - 4 )$ confines. This is in accord with the electric
theory: when $ X {\cal L} \ne \Lambda_1^4$, eq. (\ref{w1n1}) shows that
 the fields $V_{i \lamd}$ are forced to have
nonzero expectation values and break $SU(2)_2$. In the dual theory, we
can integrate out
the massive quarks. On   adding the confining superpotential generated by
$SP( 2 m - 4)$ with the $2 m$ doublets $r^a$, it is easy to see that
all $SP( 2 m - 4)$ mesons, along with the singlets ${\cal R}_{ab}$
are massive.
The massless degrees of freedom
 are the mesons $Y_{i a}$, $X$ and $\cal L$. The
superpotential for the massless degrees of freedom
vanishes. In particular, there is a point 
 in the moduli
space where all the light fields $Y_{i a}$, $X$, $\cal L$
have zero expectation values.  At that point,
all the global symmetries are unbroken and  the
theory therefore exhibits confinement without chiral symmetry breaking.
 One can  also show, as a consistency check,  that the 't Hooft anomalies are
saturated by  the light mesons at that point.

In contrast,  along
 the flat direction $X {\cal L} = \Lambda_1^4$,
the theory is in the non-Abelian Coulomb phase.

Finally, we  give  expectation values to all
electric mesons  $Y_{i a}$, $X$, $\cal L$ and ${\cal R}_{ab}$.  This makes all
dual quarks massive and one can now integrate them out thereby obtaining
a pure $SP(2m-4)$ theory in the infra-red.
The scale of this theory is given by (\ref{scalematchingmass})
  $\bar{\Lambda}_{2L}^{3 m - 3} = {\rm Pf} \tilde{M} ~
\bar{\Lambda}_2^{2 m - 4}$, where $\tilde{M}$ is the mass matrix of the
$SP(2 m - 4)$
quarks, which can be read off eq.(\ref{w1n2}).  Gaugino condensation
 in
the low energy theory generates the  superpotential  \cite{sp}
$W = (m - 1)~ 2^{m - 3 \over m - 1} ~\epsilon_{m - 1}~ \bar{\Lambda}_{2L}^3$,
where $\epsilon_{k}$ denotes the $k$-th
root of unity.
Evaluating the Pfaffian of the mass matrix $\tilde{M}$ we obtain for this
 superpotential, after using the scale matching
conditions (\ref{fsm1}),
(\ref{n1scalematching}):
\beq
\label{w1ngenericflat}
W  = ( 1 - m ) ~ \epsilon_{m - 1}~ \left[
{{\rm Pf} ~{\cal R} \over  \Lambda_2^{5 - m}}~
( X - {\Lambda_1^4 \over {\cal L}} )
- { {\cal R}^{m - 1} \cdot Y^2 \over ~2^{m+1}~ (m - 1)! ~
\Lambda_2^{5 - m}~ {\cal L}}  ~
 \right]^{1 \over m - 1}   ~,
\eeq
where all indices are contracted with
the appropriate $\varepsilon$-symbols.
We can add mass perturbations, $\delta W =
{1 \over 2} m_{\cal R} \cdot {\cal R} +
m_{\cal L} {\cal L} + m_{X} X$ to (\ref{w1ngenericflat}) and compute the
vevs of the meson fields:
\beqa
\label{exactvevsmn}
\langle X \rangle &=&
\epsilon_1
{}~\sqrt{  \Lambda_1^{4} ~ m_{\cal L} \over  m_X} +
\epsilon_2
{}~\sqrt{  \Lambda_2^{5 - m} ~ {\rm Pf} ~ m_{\cal R} \over  m_X}
 \nonumber \\
\langle {\cal L} \rangle &=&
{}~ \epsilon_1
{}~\sqrt{ m_X~ \Lambda_1^{4} \over m_{\cal L} } \\
\langle {\cal R}_{ab} \rangle &=&   ( m_{\cal R}^{-1} )_{ab}
{}~ \epsilon_2
{}~\sqrt{ m_X ~\Lambda_2^{5 - m}~ {\rm Pf}~ m_{\cal R} }    \nonumber ~,
\eeqa
where $\epsilon_1, \epsilon_2 = \pm 1$.
These vevs
 coincide with the ones
 determined by holomorphy and the 
various limits\footnote{For 
general $n, m$, with mass terms added for all fields,
the gaugino condensates of the low-energy
pure $SU(2)$ theories are determined by  the
symmetries:
\beqa
\label{gaugino}
\langle \lam_1 \lam_1 \rangle &=&
\epsilon_1 ~ 32 ~\pi^2~\sqrt{ m_X~ \Lambda_1^{5 - n} ~{\rm Pf}~ m_{\cal L} }
{}~f_1 (t) \\
\langle \lam_2 \lam_2 \rangle  &=&
\epsilon_2 ~ 32 ~\pi^2~\sqrt{ m_X ~\Lambda_2^{5 - m}~ {\rm Pf}~ m_{\cal R} }
{}~f_2 (t)
\nonumber ~,
\eeqa
where $\epsilon_{1,2} = \pm 1$, and $f_{1,2}$ are arbitrary functions
of
$t = ( \Lambda_1^{5 - n}~ {\rm Pf}~ m_{\cal L} )/
( \Lambda_2^{5 - m}~ {\rm Pf}~ m_{\cal R} )$.
As in ref.\cite{ILS}, holomorphy,
 the large mass or small $\Lambda_{1,2}$
limits, and  scale matching,
 allow us to conclude that
$f_{1,2} \equiv 1$.
{}From the Konishi anomaly equations, we can now determine the exact dependence
of the vacuum expectation values on the masses and Wilsonian gauge couplings,
which, for $n = 1$
coincide with the ones obtained from the superpotential
(\ref{w1ngenericflat}).}.
Taking the limit
of vanishing masses
in various orders now allows us to explore the moduli space. One finds from
eq. (\ref{gaugino}) that
 a generic flat direction is given by arbitrary expectation values for
$X$ and $\cal L$, while the  rank of $ {\cal R}_{ab}$ is restricted to be
$\le 2$.  We saw in our discussion of the $[n,m]$ models, in a different way,
 how this  restriction on the rank of ${\cal R}_{ab}$ arose.
Here we see, once again,  that while in the electric theory this restriction
arose classically, in the dual it arises as a consequence
of a non-perturbative effect.
The description of the flat directions obtained above in the dual theory
agrees completely
with that of the electric theory.

This brings us to the end of our discussion for the $SP(2m-4)$ dual theory.
We turn next to another dual of  the $[1,m]$ theory. It can be
obtained by first dualizing $SU(2)_2$ and subsequently dualizing the
first group. The resulting theory has an
 $SP(4m-8) \times SP(2m-4)$ gauge symmetry.
The analysis is considerably more complicated in this case and we will
be able to  carry it out only partially for the case $m=3$. Even
so, as we will  see, this constitutes a very non-trivial test of duality,
especially in view of the  quantum deformed moduli space in the electric
theory.

The matter content and superpotential of the $SP(4m-8) \times SP(2m-4)$
 theory can be
deduced from Table 3 and the accompanying discussion, in particular,
eq. (\ref{finalspotential}),
  after the following replacements have been made:
$n \rightarrow m$,
$m \rightarrow 1$, $\scL \rightarrow \scR$, $\scR \rightarrow \scL$,
 $a \rightarrow i = 1,2$ and $i \rightarrow a = 1,...,2m$.
Note that  $\mu_{1}$, $\Lambda_1$ and $\bar{\Lambda}_1$ now  refer to the {\it
second} group in Table 3 -- $SP(4 n + 2 m - 10)$, with $n \rightarrow m$ and
$m \rightarrow 1$ --
 while $\mu_{2}$, $\Lambda_2$ and $\bar{\Lambda}_2$ refer to the {\it
first} group -- $SP(2 n - 4)$, with $n \rightarrow m$ of Table 3.
We will denote, as in Table 3, by  $\lam $
the indices under $SP(2 m - 4)$, and with
$ \lamd $ -- the $SP(4 m - 8)$ ones.

We will only analyze the dynamics for the simplest 
case\footnote{Analyzing e.g. the $m = 4$  case requires understanding the
nonperturbative dynamics of $SP(4)$ with a traceless antisymmetric
tensor and ten fundamentals and superpotential given in
(\ref{finalspotential}).
For a superpotential of this type,
this is an interesting unsolved problem. We note only that the
deconfining method of \cite{berkooz} (see also \cite{p1}, \cite{luty})
is not of immediate help in this case -- the antisymmetric tensor reappears
after dualizing once.}, $m = 3$.
 Then the dual is $ SP(4) \times  SU(2)$.
The $SU(2)$ gauge group has a matter content of six doublets,
 ${1 \over \mu_1} \G{\lam}i$,
 $J_{\lam \lam_1} p_{\lamd}^{\lam_1}$, and is therefore in the confining phase.
The $SU(2)$ mesons are ${\cal M} = {1 \over \mu_1^2} G_1 \cdot G_2$,
 ${\cal N}_{i \lamd} = {1 \over \mu_1} G_i \cdot p_{\lamd} $ and
 ${\cal K}_{\lamd \lamd^\prime} = - p_{\lamd} \cdot p_{\lamd^\prime}$.
Rewriting the superpotential (\ref{finalspotential}) in terms of these mesons,
and
adding the nonperturbative piece generated by the confining $SU(2)$,
 the superpotential becomes:
\beqa
\label{w1n3}
W &=&
{-1 \over 4 ~ \mu_2 ~\mu_1^2}~\Rsc{a}{b}  ~v^{a} \cdot {\cal K} \cdot v^{b}
+ {1\over 4~ \mu_1} {\cal L} ~r_i \cdot r^i
+ {1 \over 2~\mu_1 ~\mu_2} ~Y_{a i}~ v^a \cdot r^i +
 {1 \over 2} ~ {\cal N}_i \cdot r^i \nonumber \\
&+&
{1\over 4~ \mu_1~ \mu_2^2}~ W_{a b} ~v^a \cdot v^b -
{1 \over \bar{\Lambda}_{2}^3 }~\left( {\cal M}~ {\rm Pf} {\cal K}
- {1 \over 4} ~ {\cal N}_{i} \cdot {\cal N}^i
 \cdot {\cal K}  \right) ~,
\eeqa
where $v^{a} \cdot {\cal K} \cdot v^{b} = v^{a}_{\nud}
 ~J^{\nud \lamd} ~{\cal K}_{\lamd \dot{\mu}} ~J^{\dot{\mu} \dot{\rho}}
  ~v^{b}_{\dot{\rho}}$ and
$ {\cal N}_i \cdot {\cal N}^i
  \cdot  {\cal K} =
{\cal N}_{i \lamd} ~ {\cal N}_{j  \lamd^\prime} ~ \varepsilon^{ij}
{}~{\cal K}_{\nud \nud^\prime}~ \varepsilon^{\lamd \lamd^\prime
 \nud \nud^\prime}$.
The
$SU(2)$ mesons ${\cal N}^i_{ \lamd}$ and the  quarks $r^i_{\lamd}$
are  massive.
To find the masses of the heavy fields, it is convenient to decompose
${\cal K}_{\nud_1 \nud_2} = J_{\nud_1 \nud_2} \mu_1 X/(2 \mu_2) + 
\tilde{\cal K}_{\nud_1 \nud_2}$,
using   (\ref{eom}) with  $\mu_1$ and $\mu_2$ interchanged. We denote by
 $\tilde{\cal K}$ the traceless part of $\cal K$.
The equation of motion
for the meson $\cal M$ implies
\beq
\label{reltn1}
0 = {\rm Pf} {\cal K} = { \mu_1^2 X^2 \over 4~ \mu_2^2} + 
{\rm Pf} \tilde{\cal K}~.
\eeq
The
mass
matrix of the fields $(r^i_{\lamd}, {\cal N}^i_{ \lamd})$
can be read off eq.(\ref{w1n3})  and its Pfaffian is proportional to:
\smallskip

$
{\rm det} \left( \begin{array}{cc}
                   {\scL \over   \mu_1} ~J^{\lamd_1 \lamd_2} & J^{\lamd_1
\lamd_2}      \\
                  J^{\lamd_1 \lamd_2}       &
- { \mu_1 ~X \over  \mu_2 ~ \bar{\Lambda}_2^3} ~ J^{ \lamd_1 \lamd_2} -
{2 \over\bar{\Lambda}_2^3}~ \tilde{\cal K}^{ \lamd_1 \lamd_2}  \end{array}
\right)
$
\beq
\label{pfN}
{}~ = ~ \left(~4 ~{\rm Pf}  \tilde{\cal K} ~\left({\scL \over \mu_1~ \bar
                {\Lambda}_2^3
}\right)^2 +
\left( 1 + { \scL ~  X \over \mu_2 ~ \bar{\Lambda}_2^3} \right)^2 ~\right)^2
{}~ = ~ \left( 1 + {2~ \scL ~  X \over c(3,1) ~ \Lambda_1^4} \right)^2~.
\eeq
In order to obtain the second equality in
(\ref{pfN}), we used (\ref{reltn1}),  and the scale matching relation for the
scale $\bar{\Lambda}_2$,
eq.(\ref{lambdab1}) with the appropriate replacements discussed earlier
(i.e., $\bar{\Lambda}_2^3 = c(3, 1) \Lambda_1^4/\mu_2$; 
to avoid confusion we have not interchanged $m$ and
$n$ in $c(n,m)$).
The mass matrix (\ref{pfN})  is therefore
non-degenerate, provided $X ~{\cal L} \ne \Lambda_1^4$.
The requirement that the dual theory reproduces exactly the modification of the
quantum moduli space in the electric theory thus
fixes the constant
\beq
\label{cof1fixed}
c(3, 1) = - 2~,
\eeq
which, using the recursion relations (\ref{cmrecursion}),  (\ref{cmrecursion2})
allows us to determine
the constant
in the scale matching relation  (\ref{lambdab1}).
In this case
we can integrate out the fields $\cal N$, $r$. We are left with an
$SP(4)$ theory with a traceless antisymmetric tensor $\tilde{\cal K}$
(the traceless
part
 of $\cal K$) and six fundamentals (the fields $v^a$) and a complicated
superpotential, whose precise form can be  obtained but is not essential
to the subsequent discussion. It can be shown that this theory is in
the confining phase and generates a
nonperturbative superpotential:
\beq
\label{sp4w}
 W_{n.p.} \sim v^6 ~ \tilde{\cal K}^2~.
\eeq
The subsequent algebra is straightforward but somewhat tedious
and we only describe it in words here.
Adding (\ref{sp4w}) to (\ref{w1n3}) (after integrating
out the heavy fields $\cal N$ and $r$), and rewriting it in terms of
$SP(4)$ mesons gives us the required superpotential. From it one
can see
 that all mesons gain mass, mixing with the singlets. The
equations of motion for the singlets require that the expectation
values
of the heavy meson fields vanish (hence the precise form of the superpotential
(\ref{sp4w})
 is not essential, since it vanishes after the heavy
mesons are integrated out).
 The only massless degrees of freedom that remain, finally,
are $X$, $Y_{i a}$ and $\cal L$ and the superpotential for them vanishes. 
At the origin -- where all these light fields
have zero expectation values -- one again sees that
the global symmetries are unbroken and that 
't Hooft's conditions are saturated by these fields, exactly as in the
case of the electric theory (and the $SP(2m-4)$ dual discussed above).

On the other hand, along the flat direction $X ~{\cal L} = \Lambda_1^4$,
as follows from (\ref{pfN}),
the mass matrix is
degenerate,
and the massless spectrum of the dual $SP(4)$ theory now has eight fundamentals
and
a traceless antisymmetric tensor. By using the symmetries, we can deduce that
no superpotential (which is nonsingular at the origin)
can be written in terms
of the mesons and baryons. Hence, new massless degrees of freedom have to
descend in the low-energy theory, and the theory is probably in
the non-Abelian Coulomb phase. However, as was  remarked earlier (see
footnote), presently  we do not have a sufficient understanding of the
non-Abelian
Coulomb phase of the $SP$-theories with antisymmetric tensor and fundamental
matter content to carry this analysis further. Even so, the
agreement obtained so far is already a  non-trivial check of the
equivalence between this dual and the  electric theory.

\mysection{The Confining Models.}

\label{theconfiningmodels}

We end our study   of the $SU(2) \times SU(2)$ theories by 
  considering the confining models.
Some of these  models, namely the  $[0,0], [1,0]$ and $[2,0]$, 
were studied in
ref. \cite{ILS}. In this section we study the remaining
$[2,1]$ and $[1,1]$ models (we note
that the latter were also studied in ref.\cite{ken}). 
We will see how the  exact superpotential can be
 determined in these cases.

We first consider  the $[2,1]$ model. It is convenient, in determining the
exact superpotential,  to
 consider the theory in the two limits  $\Lambda_1 \gg \Lambda_2$
and   $\Lambda_1 \ll \Lambda_2$.  We start  with the limit
$\Lambda_1 \gg \Lambda_2$. In this limit $SU(2)_1$ has 6 doublets and
 its non-perturbative dynamics   generates a superpotential:

\beq
\label{w211}
W = - {1 \over \Lambda_1^3 } ~\left(
X ~{\rm Pf} \scL - {1\over 4}~ V \cdot V \cdot  \scL \right)~,
\eeq
with $ V \cdot V \cdot  \scL =  \scL_{ij}
\V{\ald}k \V{\bed}l \varepsilon^{ijkl} \varepsilon^{\ald \bed}$.
 The
low-energy $SU(2)_2$ now has $6$ doublets, $R_a$ and ${1 \over
\Lambda_1} V_i$, and one expects a  nonperturbative
superpotential to be generated in this theory as well.  Eq.(\ref{w211})
can be viewed as giving rise to Yukawa couplings  in this low-energy
theory.
The simplest guess to make  for the $SU(2)_2$ non-perturbative
contribution is  to assume that it is unaffected by the presence of these
Yukawa couplings.  The exact superpotential  would  then be given by:
\beq
\label{w212}
W = - { X ~{\rm Pf} \scL - {1\over 4}~ W \cdot \scL
\over \Lambda_1^3} - {  {\cal R}~ {\rm Pf} W  - {1 \over 4}
 ~ Y^2 \cdot W \over \Lambda_2^4
{}~\Lambda_1^3} ~,
\eeq
where $Y^2 \cdot W = Y_{ia} Y_{jb} W_{kl} \varepsilon^{ijkl}
 \varepsilon^{ab}$ and
$W_{ij} = V_i \cdot V_j$.

This simple guess in fact turns out to be correct. There are several ways to
see this.  First, one can add masses for all the fields and show that the
resulting expectation values agree with those determined by the
Konishi anomaly.  Second, one can
add masses for a few fields and flow to the models analyzed in
ref. \cite{ILS}. One finds that eq. (\ref{w212}) correctly reduces to the
superpotential for these models.  Third, there are other terms consistent
with the symmetries that can be written besides those above, for example
one can take the ratio of any two terms in eq.( \ref{w212}) and obtain
additional terms. However, such terms always result in  singularities at
points in field space where there is no physical reason to expect  them.
 This is especially true from the point of view of the dual theory
which in this case is  a weakly coupled and completely Higgsed theory, so
that   no such
singularities can occur in it.  Finally, as we see below, proceeding in a
similar
way, we obtain exactly the same infra-red physics in the opposite limit,
$\Lambda_2 \gg \Lambda_1$.  This strongly suggests that no terms of
order $\Lambda_2/\Lambda_1$  are  being  left  out  in eq.(\ref{w212}).

In the  $\Lambda_2 \gg \Lambda_1$
limit,   $SU(2)_2$ gets strong first and its dynamics generates
the constraint  (\ref{w1n1}).
Below the scale $\Lambda_2$, a dilaton is dynamically generated,
as discussed in section 3.2, and the
Wilsonian coupling of $SU(2)_1$ is field-dependent and is determined by
eq.(\ref{fsm1}) (with $m=1$,  and $\Lambda_1$ and $\Lambda_2$
interchanged). The low-energy $SU(2)_1$  now has $6$ doublets. 
Proceeding as above, by   adding the
two superpotentials and solving the constraint for the $SU(2)_2$ meson
$V_1 \cdot V_2$, we find the following superpotential in this limit:
\beq
\label{w213}
W = - {{\rm Pf} \scL \over \Lambda_1^3 }~\left( X - {\Lambda_2^4 \over
{\cal R}} \right) + { Y^2 \cdot \scL \over 4~ \Lambda_1^3 ~\scR}~.
\eeq
Although the two superpotentials, 
eq. (\ref{w213}) and (\ref{w212}), look different they
describe the same infrared physics. For example, (\ref{w213}) can be obtained
from (\ref{w212}) by integrating out the field $W$ which is massive along the
flat direction with $\scR \ne 0$. It can be shown that along other flat
directions as well, (\ref{w213}) and (\ref{w212}) lead to the same massless
spectrum and interactions.
By adding mass terms and taking them to zero in various limits
 one can show that the moduli spaces are also identical,
as eq. (\ref{w213}) too  yields the correct vevs, given by  (\ref{gaugino})
and the Konishi anomaly. In particular, the
point of vanishing expectation values and unbroken global symmetries, which
is obviously part of the moduli space of (\ref{w212}) is also part of
the moduli space in the $\Lambda_2 \gg \Lambda_1$ limit, eq. (\ref{w213}).
At this point the theory exhibits  confinement
without chiral symmetry breaking  and the required 't Hooft
anomaly matching conditions are saturated by
the fields $X$, $\scR$ and $Y_{ia}$.

In view of all this evidence we conclude that eq.(\ref{w212}) 
(or equivalently eq.(\ref{w213})) is the exact superpotential for the low-energy $[2,1]$
theory.  As we saw above, this superpotential can be obtained
by simply summing the contributions of the two groups together.
We expect  this to be true, generally, for product group theories
when they have a    quantum moduli space; their exact superpotentials
should  therefore be straightforward to determine.

It is worth emphasizing that even though the superpotential,
eq.(\ref{w212}) (or equivalently  eq. (\ref{w213})) and the
related description in
terms of the moduli fields was derived initially in the limit when
one of the gauge couplings was much stronger than the other,
the evidence discussed above shows that it is valid more generally.

Finally, we discuss the $[1,1]$ model.
The exact superpotential in this case can be determined by integrating
out one flavor of the $L$ fields.  Starting with eq. (\ref{w212})
and integrating out in addition the heavy field $W_{12}$ we arrive at the
exact superpotential for the $[1,1]$ model (this form was obtained
previously by the ``integrating-in" procedure in ref.\cite{ken}):
\beq
\label{w112}
W = L ~\left( \scL~ \scR~ X - \scL ~\Lambda_2^4   - \scR ~\Lambda_1^4
- Y^2 \right)~.
\eeq
where $L$ is the Lagrange multiplier\footnote{
The same form is also obtained by starting with eq.(\ref{w213}) after
suitably redefining the Lagrange multiplier.}. A few comments are in
order  about this superpotential:

{\it i}. It can also be
derived by considering the theory in
 the two limits $\Lambda_1 \gg \Lambda_2$ and
$\Lambda_1 \ll  \Lambda_2$  and
adding the contributions of the two groups, as was done for the $[2,1]$
models.

{\it ii}. As expected, it  is symmetric under exchanging the two groups.

{\it iii}.  As eq.(\ref{w112})  shows, the $[1,1]$ model has a
quantum-deformed moduli space.  It is interesting to note, that although
each gauge group individually  has a quantum-modified moduli space, and
the origin is excluded from its quantum moduli space, in the product group
theory it is possible to go to the origin\footnote{By the origin we mean
the point where all the light fields have zero vevs and the global
symmetries are unbroken.  A heavy field like $W_{12}$ which is
a singlet under all  global symmetries has  a vacuum expectation value
at this point though, due
to the quantum-modified constraint.}.
At this point the global symmetries are unbroken, and 't Hooft's
 conditions are saturated by $X$, $Y_{ia}$ and $\scL$ (or equivalently,
$\scR$, since it has the same symmetries).

This concludes our discussion of the confining models.

\mysection{The ${\bf SU(N) \times SU(M)}$ Models.}

\label{thesuntimessummodels}

In this section we consider a generalization of the $SU(2) \times SU(2) $
theory, with an $SU(N) \times SU(M)$ gauge group and  the matter
content of a single field transforming under both
gauge groups, and  additional fields transforming under the first
or second
group alone as fundamentals or antifundamentals.  $N_F$ will
denote the number of flavors of the $SU(N)$ group, when $SU(M)$ is
turned off. Likewise, $M_F$ will denote the total number of flavors
under the $SU(M)$ group.  The particle content of the
model, as well as the charges of the fields with respect to the nonanomalous
$U(1)$ symmetries are given in Table 4.  Note that  in
general these theories are  chiral.

By varying the four
parameters, $N, N_f, M$ and $M_f$,  one can explore the
non-perturbative dynamics of these theories.
 We will not do so   here.   Rather we will content ourselves
with discussing  three aspects of these models.  First,  we will
study the renormalization group flows in the space of the two gauge couplings
in these theories.  These flows can be analyzed in the vicinity of the two
fixed
points  obtained by turning off one or the other gauge coupling.  We will
find a simple criterion to decide when the gauge coupling, which
is initially turned off,  is a relevant perturbation. We than use
this criterion to show that  the flows are locally 
consistent with the absence of a phase
transition. 
Second,  we  will construct a dual theory by alternately dualizing
the two  gauge groups. Finally, 
we  will analyze a subset of these theories with
an $SU(N) \times SU(N-1)$ symmetry and show that they dynamically
break supersymmetry after adding suitable Yukawa terms.

\begin{table} \begin{center}
{\centerline{Table 4: The field content and nonanomalous
$U(1)$ symmetries of  the
$SU(N) \times SU(M)$ theory. }}

\smallskip

\begin{tabular}{c|c|c|c|c|c|c}
\hline \hline
$\ $ & $SU(N)$ & $SU(M)$  & $U(1)_1$ & $U(1)_2$ & $U(1)_3$ & $U(1)_R$\\
\hline
$Q$ & $\Yfund$ & $\Yfund$  & $0 $ & $0$ & $M - N_f $ & $0$ \\
$L_{i = 1,..., N_f - M}$ & \Yfund &{\bf 1}  & $N_f$       & $0$
        & $M$
&${2 (N_f - N)\over N_f - M}$\\
$\bar{L}_{I = 1,... N_f} $ &$\overline{\Yfund}$  &{\bf 1}
& $M - N_f$ & $0$          & 0
& $0$ \\
$R_{a =  1,..., M_f - N }$ &{\bf 1} &{\Yfund}                 & $0$
   &$M_f$        &${N (N_f - M)\over M_f - N}$
& ${2 ( M_f - M) \over M_f - N}$ \\
$\bar{R}_{A = 1,..., M_f}$ &{\bf 1} &  $\overline{\Yfund}$ & $0$
& $N - M_f$ & $0$&$0$\\
\hline
\end{tabular}
\end{center}
\end{table}

\subsection{Renormalization Group Flows.}

\label{renormalizationgroupflows}

In this section we  consider the RG flows in  the space of the two
gauge couplings of the $SU(N) \times  SU(M)$ theories introduced above.
 
 In $N = 1$ supersymmetric
Yang-Mills theories  there exists an exact relation
between the anomalous dimensions of the various fields $\Phi_i$
and the beta function \cite{NSVZ} :
\beq
\label{betag}
\beta(\alpha)= - {\alpha^2\over 2 \pi}~
                              {3T(G) -  \  \Sigma_i  \ T(R_i) ~(1-\gamma_i)
\over
                                1 - T(G) ~{\alpha \over 2 \pi} }.
\eeq
where $\alpha= {g^2 \over 4 \pi}$ , $T(R)~ \delta^{ab} ={\rm Tr}_R (T^aT^b)$
and $T(G) =T(R={\rm adjoint})$. $\gamma_i$  is the anomalous dimension
of the chiral superfield $\Phi_i$ and is related to the full scaling dimension
by $\gamma_i +2 = 2 d_i$.
Our subsequent discussion will rely mainly on the numerator
on the RHS in eq. (\ref{betag}), denoted by ${\rm num} ~\beta$
\beq
\label{defnumbeta}
{\rm num}~ \beta = -{\alpha^2\over 2 \pi}~
                            \left[ 3 T(G) -  \  \Sigma_i ~ T(R_i) (1-\gamma_i)
                                        \right].
\eeq
Note that
fixed points can arise when this numerator vanishes\footnote{
The numerator is proportional to the anomaly in  the R-symmetry current
that determines the scaling dimensions at the  fixed point.}.

For the $SU(N) \times SU(M)$ models with the field content of Table 4,  the
numerators of the beta functions  for the two
gauge groups\footnote{Eq. (\ref{betag}) is valid for the beta functions of the
gauge couplings in a
product-group theory as well, as is clear, e.g. from the instanton derivation
(first two papers in ref.  \cite{NSVZ}).}
 are given by:
\beqa
\label{betas}
{\rm num} ~\beta_N  &=& - {\alpha_N^2 \over 2 \pi} \left[
3 N - N_f + {M \over 2}~ \gamma_Q  + {N_f - M \over 2}~ \gamma_{L}
+ {N_f \over 2} ~\gamma_{\bar{L}}  \right] \\
{\rm num} ~\beta_M  &=& - {\alpha_M ^2\over 2 \pi} \left[
3 M - M_f + {N \over 2}~ \gamma_Q  + {M_f - N \over 2} ~\gamma_{R}
+ {M_f \over 2} ~\gamma_{\bar{R}} \right]\nonumber ~.
 \eeqa
As we show below, equations (\ref{betas}) allow us to
 deduce some nontrivial facts
about the flow diagram in these models.
There are three cases to consider:

{\it i}.  Each  gauge group is
in the interacting non-Abelian Coulomb phase in the limit when
the other gauge interaction is
turned off, i.e. the inequalities $3 N > N_f > 3 N/2$ and $3 M > M_f > 3 M/2$
hold.

{\it ii}.  One of the two groups, say  $SU(N)$,  is in the interacting
non-Abelian  Coulomb phase,  when the $SU(M)$ interaction is turned
off.  On the other hand, the $SU(M)$ group, is in the magnetically free phase
when the $SU(N)$ coupling is turned off.  That is,
 $3 N > N_f> 3N/2$  and $M+1<M_f<3M/2$.

{\it iii}.  Each group is in the magnetic free phase, in the absence of the
other
gauge coupling, i.e.,
$N+1<N_f<3N/2$ and $M+1<M_f<3M/2$.

In each case we will analyze the behavior of the RG flows  in the
neighborhood  of the two fixed points obtained by turning one
of the gauge couplings off.  We will then speculate on the simplest
global RG flows consistent with this local behavior.
We discuss the first case at some length.  The analysis in the other
cases is quite analogous and we discuss it somewhat briefly.

\subsubsection{${\bf 3 N > N_f > 3 N/2}$  and  ${\bf 3 M > M_f > 3 M/2}$.}

We start by  setting $g_M=0$ and considering the fixed
point in the resulting $SU(N)$ theory.
A great deal of
evidence,   from the large-$N$ limit \cite{bankszaks} and especially
now from  duality \cite{seiberg},  shows that
 the $SU(N)$ theory has  a non-trivial fixed
point that is attractive (i.e. along the line $g_M=0$) in the infra-red.
At this fixed point  the
theory is simply $SU(N)$ SQCD with $N_f$ flavors and the anomalous
dimensions of all $SU(N)$-(anti)fundamentals are equal,
$\gamma_Q =  \gamma_L = \gamma_{\bar{L}} \equiv  \gamma^*_{g_N^*, 0}$.
{}From the first of eqs.(\ref{betas}) we can  then calculate the anomalous
dimensions at the fixed point $(g_N^*, 0)$:
\beq
\label{gamma1}
\gamma^*_{g_N^*, 0} = - { 3 N - N_f \over N_f }~.
\eeq

In order to understand the RG flows in the neighbourhood  of  this fixed point
we need to know  in addition the beta function of the $SU(M)$ group.
In the vicinity of the
fixed point $ (g_N^*,  0)$
 the anomalous dimensions are still given by (\ref{gamma1}), up
to small corrections proportional to the deviation from the fixed point.
Substituting the
anomalous dimension of the field $Q$ in the
second of eqs.(\ref{betas}), we obtain for the numerator of the beta
function of the $SU(M)$ gauge coupling
\beqa
\label{betas2}
{\rm num} ~ \beta_M \bigg\vert_{(g_N \simeq g_N^*, g_M \ll 1)} &=&
- {\alpha_M ^2\over 2 \pi} ~
\left[ 3 M - M_f + {N \over 2} ~ \gamma^*_{g_N^*, 0} + . . .~ \right] \\
&\simeq& {\alpha_M \over 2 \pi}
 \left[ {N \over 2~ N_f}
{}~( 3 N - N_f ) - 3 M + M_f \right] \nonumber ~.
\eeqa
It follows that the coupling $g_M$ is irrelevant if
\beq
\label{irrelevant1}
 3 M - M_f <  {N  \over 2~ N_f} ~ ( 3 N - N_f)~,
\eeq
and relevant if  the inequality is reversed\footnote{
In the case of an equality in (\ref{irrelevant1}),
we cannot draw any rigorous conclusion about the sign of the
beta function, without additional information about the terms
denoted by the ellipses in eq. (\ref{betas2}).}.
With this information in hand the RG flows are completely determined
in the vicinity of the $(g^*_N,0)$ fixed point.

We now turn to the  $(0,g^*_M)$ fixed point. Since the fixed point for
the pure $SU(M)$ theory is attractive  in the infra-red we know that  the
$g_M$ coupling is irrelevant  at this fixed point.  On considering
the effects of the $SU(N)$ group in the neighbourhood of this  fixed point we
have, from an analysis very similar to that above, that the $g_N$ coupling is
irrelevant provided
\beq
\label{irrelevant2}
 3 N - N_f < {M \over 2~ M_f}~ ( 3 M - M_f)  ~,
\eeq
and is relevant if this inequality is reversed.

We find   four possibilities depending on whether  the
two inequalities, (\ref{irrelevant1}) and (\ref{irrelevant2}) are met or not.
We discuss these in turn.

First, we note that (\ref{irrelevant1}) and (\ref{irrelevant2})
cannot be simultaneously satisfied:
iterating them once we obtain
the inequality $3 M - M_f \le { N \over 2 N_f}
{M \over 2 M_f } (3 M - M_f)$, which is
clearly violated, since $ { N \over 2 N_f}$ and
${M \over  2 M_f }$  are both smaller than  $1\over 3$ (recall that we are
considering the theories in the conformal window).
This rules out one possibility. Had it been allowed,  there would have
to be a phase boundary in the space of the two couplings and associated
with it a phase transition as the two couplings are varied.
For example, the simplest global flow diagram consistent with this
possibility is shown in Fig. 1(c). One sees that for $g_N$ small
enough the theory flows to the $g_N=0$ fixed point, in contrast when $g_N$ is
big enough it flows to the $g_M=0$ fixed point.

There is some lore \cite{seibergwitten}, \cite{phasesofn1theories}, that
phase transitions are not allowed in SUSY theories.
As best as we can tell, this lore should apply
to couplings in the superpotential.
For such couplings holomorphy suggests that a phase boundary would have to
be of (real) codimension two. But a surface of codimension two is
not a boundary at all, since one can interpolate around it. Thus  there  are no
phase boundaries and therefore no phase transitions can occur. It is not clear
to us, if this lore is directly applicable in the present case. We note for
example, that the gauge couplings under consideration are {\it not} the
Wilsonian gauge
couplings and their renormalizations involve the anomalous dimensions (eq.
(\ref{betag}))
which are not holomorphic functions of the gauge couplings.
Nevertheless,  our results in this case are consistent with the absence of
a phase transition. In fact,   the subsequent discussion  in this section
will be   consistent with the absence of phase
transitions too.  It is worth noting that this was  equally true for   our
discussion  of $SU(2) \times SU(2)$ duality.

\begin{figure}
\psfig{file=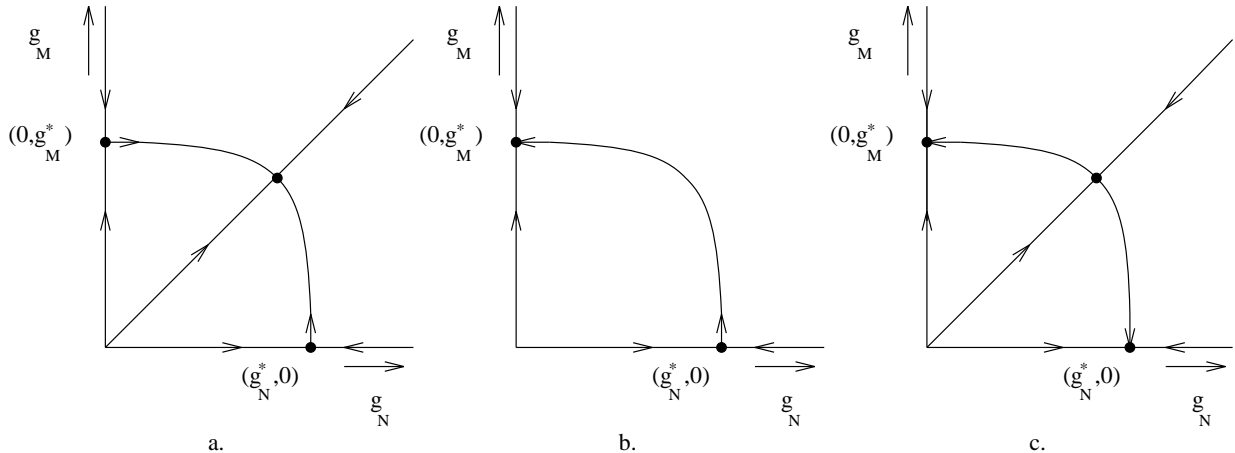}
\caption{Three different RG flow diagrams in the space of the two
gauge couplings.  (a) and (b) are the simplest flows consistent with the
behavior in the vicinity of the fixed points, $(g_N^*,0) $ and $(0,g_M^*)$,
while (c) is not realized (see text).}
\end{figure}

We now return to considering the remaining possibilities presented
by (\ref{irrelevant1}) and (\ref{irrelevant2}).
If none of these inequalities  hold, then at the fixed point for each
gauge group the other gauge coupling is relevant (fig. 1(a)),
and we expect the theory to flow to a
nontrivial fixed point where both couplings are  nonvanishing.

One would like to understand the global nature of these RG flows
 somewhat better. One limit in which
they can be analyzed is
 the large-$N$ limit first used for this purpose by
Banks and Zaks \cite{bankszaks}. In this limit,
$N, N_f, M, M_f \rightarrow
\infty$, with  $M g_M^2$ and $N g_N^2$
 kept fixed,  and $N_f/N = 3 - \epsilon_N$ and $M_f/M= 3- \epsilon_M$.
One finds, in this case, that the flow diagram can be explicitly worked out
in the region where $g_N^2 N$ and $g_M^2 M$ are both small.
The resulting flow diagram is shown in Fig. 1(a).
There is a unique
fixed point (away from both the axes) and it is infra-red attractive with
respect to both the gauge couplings.

Aside from the large-$N$ limit one cannot in general make any rigorous
statement about the global nature of the RG flows. Nevertheless
it is worth noting that the simplest RG flow diagram consistent
with the local analysis performed above continues to correspond to
Fig 1(a).
In fact  this simple ansatz for the flow diagram
is also the only one consistent with the absence of phase transitions.
If there were another fixed point there would also necessarily be
a phase boundary and therefore a possible    phase transition.

Returning to the inequalities (\ref{irrelevant1})  and (\ref{irrelevant2})
we find there is one more possibility. Namely that only one of the
two, (\ref{irrelevant1})  or (\ref{irrelevant2})
holds\footnote{
The   region in the 4-dimensional parameter space where,
 eq.(\ref{irrelevant1}), $N_f \ge M, M_f \ge N$,
and $3 N > N_f > 3 N/2,  3 M > M_f > 3 M/2$ hold simultaneously,
is quite
complicated. However, it is easy to show that it  is not empty, by finding a
particular point, e.g. $ N = 6, N_f = 10, M = 8, M_f = 22$,
that satisfies all inequalities.}. In this case again a large-$N$ analysis
can be performed. The resulting flow diagram is shown in Fig. 1(b) 
(for definiteness this diagram is drawn for the case 
when (\ref{irrelevant2})
is met but not (\ref{irrelevant1})).
One sees that in this case the theory flows to the corresponding
Seiberg fixed point in the infra-red.
For  the general case again one cannot make any rigorous statement but
Fig. 1(b) continues to be the simplest flow diagram consistent with the
local information at hand. It is also the only
diagram without any phase transitions.

\subsubsection{${\bf 3 N > N_f> 3N/2}$ and ${\bf M+1<M_f<3M/2}$.}

In the case when one, or both, theories are in the magnetic free phase, the
flows
have to be discussed separately. First we
consider the case when one of the groups, say $SU(N)$,
is in the interacting non-Abelian Coulomb phase (i.e. $N_f > 3 N/2$),
while the other
is in the magnetic free phase ($M + 2 \le M_f \le 3 M/2$).
Turning on  weakly
the $SU(M)$ gauge coupling at the $SU(N)$ fixed point, we find
the same condition (\ref{irrelevant1}) for
$g_M$ to be irrelevant. This condition cannot be satisfied now:
note that for $ 3 N/2 < N_f < 3 N$, the r.h.s. of the inequality
 (\ref{irrelevant1}) obeys $ N (3 N - N_f)/(2 N_f) < N/2$. On the other hand,
 for our model
 $N \le M_f$  (see Table 4). Therefore, from (\ref{irrelevant1}) we
obtain that $3 M - M_f < M_f/2$,  or $ M_f > 2 M$, which contradicts $SU(M)$
being in the magnetic free phase. This implies that $g_M$ is always relevant at
the
$(g_N^*, 0)$ conformal fixed point. With the analysis of the
flows in the vicinity of the $(g_N^*,0)$ fixed point complete,
 we now turn to
the $(0,g_M^*)$ fixed point.
When $g_N = 0$, the theory
flows to a magnetic
free phase. The gauge group is $SU(M_f - M)$ and the degrees of freedom are
the gauge singlet
mesons $Q \cdot \bar{R}$ and $R \cdot \bar{R}$ and the dual quarks.
Weakly gauging  the
flavor subgroup $SU(N) \subset SU(M_f)_L$, we find that it has
$N_f  + M_f  - M$ flavors,
i.e. $M_f - M$ more flavors than at the ultraviolet fixed point. Its
beta function near the infra-red free magnetic fixed point is approximately given by
\beq
\label{magn1}
\beta_N  \simeq - {\alpha_N^2 \over 2 \pi} ~
\left[ ~ 3 N - ( M_f + N_f - M )~\right]~.
\eeq
Therefore, the coupling $g_N$ is irrelevant or relevant at this
fixed point depending on whether the following inequality is met or not:
\beq
\label{irrelevant3}
3 N -  N_f < M_f - M~.
\eeq

This presents us with two possibilities, both of which can
be realized (in terms of  the allowed values of $N,N_f,M$ and $M_f$).
If (\ref{irrelevant3}) is met, the simplest ansatz for the RG flows would be
that the theory flows to the magnetic free phase of
$SU(M)$ in the infrared,
with the $SU(N)$
gauge coupling becoming irrelevant
(i.e.  the flow of Fig. 1(b))\footnote{An explicit solution is
e.g. $N = 26, N_f = 77, M = 22, M_f = 24$. Note that $SU(N)$ is
close to becoming infrared
free, $N_f = 3 N -1$; gauging the $SU(M)$ flavor symmetry is enough to drive
$g_
N$
infrared free.}.  If the inequality (\ref{irrelevant3}) is reversed, then
the simplest  RG flow diagram would correspond
 to having  one  non-trivial
fixed point away from both axes,  and
attractive in the infra-red as in Fig. 1(a).

\subsubsection{${\bf N + 2 \le N_f \le 3 N/2}$ and ${\bf M + 2 \le M_f \le 3
M/2}$ .}

Finally we can consider the case when both groups are in the magnetic free
phase, i.e. $N + 2 \le N_f \le 3 N/2$ and $M + 2 \le M_f \le 3 M/2$ hold.
In this case, the beta function for $g_N$ is still
given by eq. (\ref{magn1}), in the vicinity of the $(0, g_M^*)$ fixed
point. In an analogous manner analyzing the $g_M$ coupling in the vicinity of
the $(g_N^*,0)$ coupling gives the beta function for $g_M$ as:
\beq
\label{magn2}
\beta_M  \simeq - {\alpha_M^2 \over 2 \pi} ~
\left[ ~ 3 M - ( N_f + M_f - N )~\right]~.
\eeq
Thus $g_M$ is irrelevant if  the
\beq
\label{irrelevant4}
3 M - M_f < N_f - N
\eeq
condition is met.

However in this case it is easy to show that neither
(\ref{irrelevant3}) nor (\ref{irrelevant4})  can hold.
Eq. (\ref{irrelevant3}) implies $3 N - N_f < M_f - M \le M/2 \le N_f/2$,
which requires
$N_f > 2 N$ and contradicts $SU(N)$ being in the magnetic free phase; a similar
contradiction follows from (\ref{irrelevant4}). Therefore in this case,
we expect the theory to flow to a non-trivial fixed point. The correspondingly
simplest flow diagram in this case is given by Fig. 1(a).

\subsection{Duality in ${\bf SU(N) \times SU(M)}$.}

\label{dualityinsuntimessum}

In this section we briefly discuss how dual theories for the
$SU(N) \times SU(M)$ models can be constructed by alternately
dualizing the two groups. The analysis  is very close
to that of the  $SU(2) \times SU(2)$ theories
which were discussed at length in Section \ref{thenmmodels}.
Consequently we do not
describe the construction of the duals in detail and mainly present the
 resulting final form.

The field content of the $SU(N) \times SU(M)$ theory was
described in Table 4.  As was discussed in Section \ref{thenmmodels}
two kinds of duals can be constructed: those in which one
of the groups is dualized and those in which both groups
are dualized.  Here we only consider the duals obtained by dualizing
both the $SU(N)$ and $SU(M)$ groups. For definiteness we describe  the dual
obtained by dualizing   the $SU(N)$ group followed by  the
$SU(M)$ group below.  This  dual theory has an 
$SU(N_f-N) \times SU(M_f+N_f-N-M)$ gauge symmetry. The matter
content consists of  fields  transforming under the dual gauge group as
shown in Table 5 (we explicitly display the nonabelian global symmetries
indices; they have the same ranges as
in Table 4, and lower (upper) indices denote fundamental
(antifundamental) representations).

\begin{table} \begin{center}
{\centerline{Table 5: The field content of the dual of $SU(N) \times SU(M)$.}}
\label{table33}
\begin{tabular}{c|c|c} \hline \hline
$\ $ &$SU(N_f - N) $& $SU(M_f + N_f $ \\
$\ $ & $\ $ & $ ~ ~ - N - M)$  \\
\hline
$ \bar{p} $ & $\overline{\Yfund}$ & $\overline{\Yfund}$  \\
$l^i$ & \Yfund & {\bf 1}\\
$r^a$ & {\bf 1} & \Yfund \\
$v^I $ & {\bf 1} & \Yfund \\
${\bar r}^A$ & {\bf 1} & $\overline{\Yfund}$ \\
$G_a$ & \Yfund & {\bf 1} \\
\hline
\end{tabular}
\end{center}
\end{table}

In addition the dual theory contains the following gauge singlet fields:
\beqa
\label{singletsindual}
Y_{IA} &=&  \bar{L}_I \cdot Q \cdot \bar{R_A} \nonumber \\
{\cal R}_{aA} &=& R_a \cdot \bar{R}_A \\
{\cal L}_{iI} &=& L_i \cdot \bar{L}_I \nonumber ~,
\eeqa
whose charges   with respect to the nonanomalous
global symmetries can be determined from Table 4 and their definitions
(\ref{singletsindual}).
 The dual theory also has a superpotential given by:
\beq
\label{suWdual}
W  = - {1 \over \mu_1 ~ \mu_2} ~ {\cal L}_{iI} ~ l^i \cdot v^I \cdot
 \bar{p} +
{1\over \mu_2} ~ G_a \cdot r^a \cdot \bar{p} + {1 \over \mu_2} ~
{\cal R}_{aA}~ r^a \cdot \bar{r}^A +
{1\over \mu_1 ~\mu_2} ~Y_{IA} ~v^I \cdot \bar{r}^A~.
\eeq
The  strong coupling scales of the gauge groups in the dual theories obey
matching relations which can be derived
similarly to the corresponding relations in the
$SU(2) \times SU(2)$ theories.

The detailed analysis of these dual theories is left  for future study. We
expect that such an analysis will help in understanding
the non-perturbative behavior of
the  $SU(N) \times SU(M)$  theories, much as it did in the $SU(2) \times
SU(2)$ case.

We conclude with one comment.  The process of alternately dualizing
the groups described above  can be  continued  to construct 
other dual theories. In this case a finite set of duals of the 
$SU(N) \times SU(M)$ theories is generated.
Starting with the second dual -- the 
$SU(N_f - N) \times SU(N_f + M_f - N - M)$ theory constructed above -- 
and  dualizing the
$SU(N_f - N)$ gauge group we obtain a 
third dual: an
$SU(M_f - M) \times SU(N_f + M_f - N - M)$ gauge group with 
$N_f + M_f - M - N$
flavors of $SU(M_f - M)$ and $N_f + M_f - M$ flavors of
$SU(N_f + M_f - N - M)$. This theory has the
same gauge group, field content and superpotential 
as  the second dual of the $SU(N) \times SU(M)$ theory,
constructed by dualizing the $SU(M)$ gauge group first. 
In other words, the chain of duals that can be constructed 
by alternately dualizing the two gauge groups closes.

\subsection{${\bf SU(N) \times SU(N-1)}$ and Supersymmetry
Breaking.}

\label{suntimessun-1andsupersymmetrybreaking}

In this section we consider a subclass\footnote{The models for general $M$ are
considered
in a forthcoming publication \cite{us3}.}  of
the $SU(N) \times SU(M)$ models:
the models with $N_f = N - 1$, $M = N - 1, M_f = N$, which we will refer  to as
the
$SU(N) \times SU(N-1)$ models. We will see that these models dynamically
break supersymmetry after
suitable  Yukawa couplings are added in the superpotential.
The $SU(N) \times SU(N-1)$ models are further generalizations of the
$SU(3)\times SU(2)$ model \cite{ads} of dynamical supersymmetry
breaking. Other generalizations
have been considered in
\cite{dnns}, \cite{it}, \cite{lisa}.

We begin by analyzing the classical moduli space of these theories.
It is described, \cite{ads}, \cite{lutywati}, by
the gauge invariant
chiral superfields  $Y_{IA}$, defined in (\ref{singletsindual}), the field
$\bar{\cal B} = b^{\alpha}  (\bar{L}^{N - 1}) _{\alpha}$,
where $b^{\alpha} = (Q^{N-1})^{\alpha}$ ($\alpha$ is the $SU(N)$ index), and
$\bar{b}^A = (\bar{R}^{N-1})^A$. These fields obey the classical
constraints  $Y_{IA}~ \bar{b}^A = 0$ and $\bar{\cal B} ~\bar{b}^A  \sim
(Y^{N-1})^A$.  Adding  a  renormalizable tree-level superpotential
\beq
\label{sb9}
W_{tree} = \lambda^{IA} ~ Y_{IA}~,
\eeq
one finds that
for a Yukawa matrix of maximal rank ($N-1$)
the $Y_{IA}$ and $\bar{\cal B}$ flat directions are lifted.  To see this,
consider
 the $\bar{R}_A$ equation of motion. It  implies that
$\bar{L}_I \cdot Q_{\dot{\alpha}} = 0$ (recall that rank $\lambda  = N - 1$),
which in turn implies
that $Y_{IA} = 0$ and
$\bar{\cal B} \sim {\rm det} ( \bar{L}_I \cdot Q_{\dot{\alpha}} ) = 0$.
However, the flat directions
corresponding to
$\bar{\cal B} = 0$,  $\bar{b}^A \ne 0$ and
 $Y_{IA}=0$ are not lifted.  While the $SU(N-1)$ group is completely
broken along these directions the $SU(N)$ group is completely unbroken.
These $\bar{b}^A \ne 0$
flat directions can be lifted in the
classical theory  if in addition to the Yukawa coupling
in eq.~(\ref{sb9}) we add a term $\alpha_A \bar{b}^A$
so that the full tree-level superpotential is:
\beq
\label{sblast}
W_{tree}=\lambda^{IA} ~ Y_{IA} + \alpha_A \bar{b}^A.
\eeq
Performing an analysis similar to the one above, one can show that all
classical flat directions are lifted once
 $\alpha_A$ and $\lambda^{IA}$  are chosen such that
$\lambda^{IA} \alpha_A \ne 0$.

We turn next to the   quantum theory and first study it  in the limit
$\Lambda_1 \ll \Lambda_2$. In this limit, the
$SU(N-1)$ group is confining, with $N$ flavors, and nonperturbative
effects generate \cite{seibergexact}
a superpotential
\beq
\label{sb10}
W = { b^{\alpha} ~ M_{\alpha A}~ \bar{b}^A - {\rm det}~ M \over
\Lambda_2^{2 N - 3} },
\eeq
where the mesons of $SU(N-1)$ are $M_{\alpha A} = Q_{\alpha} \cdot \bar{R}_A$,
 and the baryons $b^{\alpha}$ and $\bar{b}^A $
were defined above. Now we weakly gauge the $SU(N)$ global symmetry.
The low-energy
$SU(N)$ gauge group has $N$ flavors, with $M_{\alpha A} \sim \Yfund$ and
$b^{\alpha}, \bar{L}_{I}^{\alpha} \sim \overline{\Yfund} ~$.
Consequently, the $SU(N)$ gauge group confines as well and dynamically
generates a superpotential. The scale of the low-energy $SU(N)$ theory is
$\Lambda_{1_L}^{2 N} = \Lambda_1^{2 N + 1}/ \Lambda_2$.
Its mesons are the fields $Y_{IA}$,
 the field ${\cal P}_A = b \cdot M_A$ (this field vanishes classically),
while the baryons are $\bar{\cal B}$
and ${\cal B} = {\rm det} M$ (which vanishes classically as well).
The exact superpotential of
this  model is then given by\footnote{This superpotential can be justified by
arguments
analogous to those of Section \ref{theconfiningmodels}. The
form of the quantum modification of the constraint  in (\ref{sb12} )
  arises after taking account of the scale $\Lambda_{1_L}$ of the
low-energy $SU(N)$ theory, as well as the fact that some of its ``quarks"
are $SU(N-1)$-composites
and are normalized with appropriate powers of $\Lambda_2$.}:
\beq
\label{sb12}
W = { {\cal P}_A ~ \bar{b}^A - {\cal B} \over \Lambda_2^{2 N - 3}}
+ A ~ \left( {\cal P} \cdot Y^{N - 1} - \bar{\cal B} ~ {\cal B} -
\Lambda_1^{2 N + 1}  \Lambda_2^{2 N - 3} \right) ~.
\eeq

We now add the tree-level perturbation $W_{tree} = \lambda^{IA} Y_{IA} + \alpha_
A \bar{b}^A$.
In order to lift all classical flat directions
 we choose,
$ \lambda^{IA} = \delta^{IA} \lambda^I $, if $ A < N$,
$ \lambda^{IN} = 0 $,
and $\alpha_A = \delta_{1A} \alpha$.
The $F$-term equations of motion  for the fields $\bar{b}^A$
that follow from this superpotential then imply that :
\beq
\label{saone}
{\cal P}_1 /{\cal P}_N \rightarrow \infty \; ,
\eeq
while the $F$-term equations for ${\cal P}_A$ and $Y_{IA}$ imply that
\beq
\label{satwo}
{\cal P}_1 /{\cal P}_N  \rightarrow 0 \; .
\eeq
In deriving eq.~(\ref{satwo}) it is useful to think in terms of the 
mesons of $SU(N)$, ${\cal{P}}_A$ and $Y_{IA}$ with masses, 
$\bar{b}^A/\Lambda_2^{2N-3}$ and  $\lambda^{IA}$ respectively. 
The vevs of these mesons can be expressed in terms of their masses
in the standard way and this leads to eq.~(\ref{satwo}) 
for ${\cal{P}}_1$ and ${\cal{P}}_N$.

 Clearly, eq.~(\ref{saone}) and eq.~(\ref{satwo})
are not compatible.  
Thus not all of the $F$-term conditions can 
 be satisfied and the theory breaks supersymmetry.

One assumption made in the discussion above  was that the
K\" ahler potential is nonsingular
in terms of the light
moduli fields  $Y_{IA}, \bar{\cal B}$ and $\bar{b}^A$.
There are two kinds of singularities one might worry about: 
those that can occur in the finite region of  moduli space and those that can
occur  at the boundaries of moduli space (when some fields go to
$\infty$).  We do not expect singularities in the finite region of moduli
space in the cases considered here.
For example, even though we do not do so, the description used above
can  be derived from a weakly-coupled, completely Higgsed, dual theory
in which no singularities can occur in the finite region of moduli space.
The finite region of moduli space can also be studied,  ref. \cite{hitoshi},
\cite{us},
by adding one more flavor of
the $L$ fields and analyzing the resulting theory which now has a
quantum moduli space and showing that the moduli fields saturate
the 't Hooft anomalies.
Finally, one does not expect singularities at the boundary of moduli space
 to be relevant to the discussion of SUSY breaking, since all flat
directions are lifted classically and we therefore do not
expect to be driven to infinite field values in the quantum theory.

It is also worth commenting on the relation between $R$ symmetries and
SUSY breaking in these models. First we consider the  choice of the tree-level
superpotential made in the discussion of SUSY breaking above; $ \lambda^{IA} =
\delta^{IA} \lambda^I $, if $ A < N$,
$ \lambda^{IN} = 0 $,
and $\alpha_A = \delta_{1A} \alpha $. In this case one finds that the
superpotential eq.~(\ref{sblast}) preserves a flavor-dependent, non-anomalous
 $R$ symmetry. The charge assignments of the fields under this symmetry are,
$\bar{R}_N \sim 2( 3 N - N^2 -1)/(N-1) $,
$\bar{R}_{A < N} \sim  2 N/(N-1) $, $Q \sim  - 2/(N - 1)$ 
and $\bar{L}_I \sim  0 $. 
 Thus, the fact that this model breaks supersymmetry, 
as we found above,
is in accordance with the
discussion of ref.~\cite{nelsonandseiberg}.  

In fact we could have used this $R$ symmetry to
argue that SUSY is broken. For this purpose note that all the flat directions
are lifted in the classical theory 
 with the superpotential of eq.~(\ref{sblast}). 
Furthermore, 
all the gauge invariant moduli, 
entering the superpotential eq.~(\ref{sb12}),
except for ${\cal P}_A$ with $A \ne N$, 
are charged under the $R$ symmetry. 
Therefore, 
if all the $F$-term conditions are met the
 $R$ symmetry must be broken. Since there are no
classical flat directions we conclude that SUSY 
is broken. 
The only other alternative is that the $F$-term constraints are not
all met but then again SUSY must be broken. 

In contrast consider the case of a tree-level superpotential with
$ \lambda^{IA} = \delta^{IA} \lambda^I $, if $ A < N$,
$ \lambda^{IN} = 0 $,
$\alpha_1$  and $\alpha_N$ $\ne 0$ and all other $\alpha_A =0$.
Since the condition $\lambda^{IA} \alpha_A \ne 0$ is met all the flat
directions are lifted classically but in this case   one can show that
there is no non-anomalous $R$ symmetry.  However, an argument similar to
 the one given above, eq.~(\ref{saone}) and eq.~(\ref{satwo}),
allows us to conclude that SUSY is broken in this case as well.

We close this section with some  
comments\footnote{We are very thankful to
M.~Dine and Y.~Shirman for  illuminating discussions in this regard.
}  
on  the case of vanishing $\alpha$.
As we saw above, in this case,  the classical flat direction
with $\bar{b}^A \ne 0$ is not lifted.
 Along this flat direction, the $SU(N - 1)$
gauge group
is completely broken, while the $SU(N)$ group is unbroken.
Thus one might expect quantum
effects to play an important role along it. 
For simplicity, we set 
$ \lambda^{IA} = \delta^{IA} \lambda^I $, if  $ A < N$ and 
$ \lambda^{IN} = 0 $ as above.
In this case, there is one solution to the $F$-term equations obtained
from the exact superpotential given by the sum of eq.~(\ref{sb12}) and
eq.~(\ref{sb9}). 
This solution involves ${\cal P}_A \rightarrow 0$,
${\bar b}^N \rightarrow \infty$ and $Y_{II} \rightarrow\infty$ 
for $I=1\ldots N-1$.
This ``runaway'' solution is somewhat surprising, since $Y_{IA}\ne 0$
is not a classically flat direction. 
To understand the origin of this solution
consider the 
 flat direction along which the fields $\bar{R}_A^{\ald}$, with 
$A <N$  have vacuum expectation values
which  we denote as  $\bar{R}_A^{\ald}=  \bar{R}
\delta_A^{\ald}$.  The flat direction then corresponds to varying $\bar{R}$.
When
${\bar R} \rightarrow \infty$, the Yukawa term in $W_{tree}$
gives a large mass to all $SU(N)$ flavors. The low-energy theory along this
classical flat
direction is then a pure $SU(N)$ gauge theory. 
Gaugino condensation in this low-energy
theory generates a superpotential
$ W \sim \Lambda_{1L}^3 =
 (\Lambda_1^{2 N + 1} ~  {\rm det} \lambda  ~  \bar{R}^{N-1 })^{1\over N}$.
The gradient of this superpotential with respect to $\bar{R}$ is proportional
to  $\bar{R}^{-{1\over N}}$, and pushes
the field $\bar{R}$ to infinity, thus
restoring supersymmetry. Strictly speaking, 
the behavior of the energy along this
direction depends on both the superpotential and the  K\" ahler potential.
Since, as mentioned above,  the  infra-red $SU(N)$ theory  is strongly coupled
along this flat direction, the K\" ahler potential is difficult to 
determine exactly.  Nevertheless, some preliminary analysis
suggests that the corrections to the classical 
K\"ahler potential for $\bar{R}$  are small 
along this direction, leading to the conclusion that SUSY is probably restored.
 
It is also worth mentioning that the behavior of the superpotential 
along the baryonic flat direction discussed above can  also  be recovered
from the exact superpotential, eq.~(\ref{sb12}). On adding the tree-level
superpotential eq.~(\ref{sb9}),   one can solve for all the other fields
through their $F$-term equations in terms of the antibaryon $\bar{b}^N$. 
This gives a superpotential $ W \sim 
 (\Lambda_1^{2 N + 1}  ~ {\rm det} \lambda  ~  \bar{b}^N )^{1\over N}$,
which is identical to the one obtained above once one identifies
$\bar{R}^{N-1}$ with ${\bar b}^N$.

\mysection{Acknowledgements.}

We would like to thank Michael  Dine,  Don Finnell, 
Lisa Randall and especially David Kutasov and Yuri Shirman for 
illuminating discussions. 
E.P. acknowledges support by a Robert R. McCormick
Fellowship and by DOE contract DF-FC02-94ER40818. Y.S. and S.T. acknowledge
the support of DOE contract DE-AC02-76CH0300.

\appendix

\mysection{Notations. Duality and Scale Matching for ${\bf SP(2N)}$.}

\label{appendixa}

We take $SP(2N)$ to denote the $SP$ group whose fundamental representation is
of
dimension $2N$ so that $SP(2)=SU(2)$. The dimension of the $SP(2N)$ group is
$N ( 2N +1)$. The generators $T^a$ for a representation $R$ obey
${\rm Tr} T^a T^b = T(R) \delta^{ab}$, where $T(\Ysymm) = N + 1$,
$T(\Yfund) = T(\overline{\Yfund}) =  1/2$ and $T(\Yasymm) = N - 1$.
Here \Yasymm $~$ is a traceless
antisymmetric
tensor, and \Ysymm $~$ is the adjoint representation (symmetric tensor).

The one-loop coefficient of the beta function of
the $SP(2N)$ theory with $2N_f$ fundamentals $Q_{i \lambda}$,
($i = 1,...,2 N_F$,  $\lambda = 1,...,2N$) is
 $b_0 = 3 T(\Ysymm) - 2 N_f  T(\Yfund)  = 3 N + 3 - N_f$.
The $D$-flat directions of the electric theory are described by
the gauge invariant
chiral superfields (the $SP(2N)$-``mesons")
\beq
\label{spmesons}
M_{ij} = Q_{i \lambda_1} ~J^{\lambda_1 \lambda_2} ~ Q_{j \lambda_2}
\equiv Q_{i} \cdot  Q_{j}~.
\eeq
Eq. (\ref{spmesons}) defines our summation convention:
we always take the $SP$-fundamentals
to have lower indices and  raise (lower) $SP$ indices
using the $SP(2N)$ invariant antisymmetric tensors
\beq
\label{Jupper}
J^{\mu \nu}  = {\rm diag} \left\{ ~\left( \begin{array}{cc} 0 & 1\\
-1 & 0 \end{array} \right),~...~,
 \left( \begin{array}{cc} 0 & 1\\ -1 & 0 \end{array} \right)   ~\right\}
\eeq
and
\beq
\label{Jlower}
J_{\mu \nu}  = {\rm diag} \left\{ \left(\begin{array}{cc} 0 & - 1\\ 1 & 0
\end{array} \right),~...~,
 \left( \begin{array}{cc} 0 &- 1\\ 1 & 0 \end{array} \right)   \right\} ~,
\eeq
such that $J^{\mu \nu} J_{\nu \lambda} = \delta^{\mu} _{\lambda}$ and
$Q^{\alpha} = J^{\alpha \beta} Q_\beta$.
For $SP(2)$ we denote the invariant tensor by
 $\varepsilon_{\alpha \beta}$  with $\varepsilon^{12} =
\varepsilon_{21} = 1$, in accord
with (\ref{Jupper}, \ref{Jlower}).

For $N_f \le N + 2$, the theory is confining \cite{sp}. The
$N_f = N + 2$ theory has a dynamically
generated superpotential:
\beq
\label{spsuperpotential}
W = -{ {\rm Pf} M \over 2^{N - 1} ~ \Lambda^{2 N + 1}}~,
\eeq
where $\Lambda$ is the scale of the $N + 2$ flavor theory.
The superpotentials for $N_f < N+2$ can be obtained from
(\ref{spsuperpotential}) by integrating out
extra flavors and using the scale matching relations
(\ref{scalematchingmass}) given below.
The Pfaffian of a $2 N_f$-dimensional antisymmetric matrix is defined as
$$
{\rm Pf} M =
 {1 \over  2^{N_f}~ N_f! }~
\varepsilon^{i_1 i_2 ... i_{2 N_f -1} i_{2 N_f}} ~ M_{i_1 i_2  } ~. . .
{}~M_{i_{2 N_f -1} i_{2 N_f }}
 \equiv {M^{N_f} \over  2^{N_f}~ N_f!  }~,
$$
and its square equals the determinant.  We use
 $\varepsilon^{1 2 3 ... (2N_f - 1)  (2N_f)} = + 1$.

For $N_f > N + 2$ the theory has a dual description
in terms of an $SP(2 N_f - 2 N - 4)$ gauge theory \cite{sp},
with $2 N_F$ fundamentals and additional gauge singlets, $M_{ij}$,
which are mapped to the
mesons of the electric theory (\ref{spmesons}), and a superpotential
\beq
\label{spdualw}
W = {1 \over 4 ~ \mu} ~ M_{ij} ~q^i \cdot q^j ~.
\eeq
The fields $q^i$ are the dual quarks and are in the antifundamental
representation of the
$SU(2 N_f)$ global flavor symmetry.  The strong coupling scale
$\Lambda$ of the electric
theory, the strong coupling scale $\bar{\Lambda}$ of the
magnetic theory and the parameter $\mu$ in (\ref{spdualw}) obey
the scale matching
relation:
\beq
\label{spscalematching}
\Lambda^{3 N + 3 - N_f } ~\bar{\Lambda}^{2 N_f - 3 N - 3 } =
16 ~ (-)^{N _f - N - 1} ~ \mu^{N_f} ~.
\eeq

Upon integrating out a flavor of $SP(2N)$ fundamentals of mass $M$,
the scale  $\Lambda_{L}$ of the low-energy  $SP(2N)$ theory with
$N_f -1$ flavors
is related to the scale of the $N_f$-flavor theory by the
$\overline{DR}$ threshold
relation \cite{finnellpouliot}:
\beq
\label{scalematchingmass}
\Lambda_L^{3 N + 3 - (N_f - 1)} = M \Lambda^{3 N + 3 - N_f } ~.
\eeq

Along a flat direction, such that, say, $M_{12} = Q_1 \cdot Q_2 = v^2 \ne 0$,
the $SP(2 N)$ theory
with $N_f$ flavors breaks to $SP(2 N - 2)$ with $N_f - 1$ flavors.
The scale $\Lambda_L$ of
the low-energy $SP(2 N - 2)$,  $N_f - 1$ flavor theory is given by the
threshold relation at the mass scale of the heavy $SP(2 N)$ vector bosons
that transform as fundamentals of the unbroken $SP(2 N - 2)$ group:
\beq
\label{scalematchingflat}
\Lambda_L^{3 (N - 1)  + 3 - (N_f - 1)}  =  {2 \over v^2} ~
\Lambda^{3 N + 3 - N_f } ~.
\eeq

\newpage

\mysection{The Anomalous Symmetries}

\label{appendixb}

Table 6 gives all anomalous $U(1)$ symmetries in the $[n,m]$ model
and its first
(Table 2) and second (Table 3) duals. $U(1)_X$ is an anomalous $R$
symmetry, under
which the superspace coordinate  $\theta_{\al} $ has charge unity.
The nonanomalous
 $U(1)$ and $U(1)_R$ from Tables 1-3 are linear combinations of
 the symmetries
in Table 6.

\begin{table}\begin{center}

{\centerline {Table~6: The anomalous $U(1)$ symmetries in the $[n,m]$
models and their duals.}}

\vspace{0.2cm}

\label{tab6}

\begin{tabular}{c|c|c|c|c|c|c} \hline\hline

\  & $U(1)_Q$ & $U(1)_1$ & $U(1)_2$ & $U(1)_X$ & $U(1)_{\mu_1}$
&$U(1)_{\mu_2}$\\
\hline
$Q_{\al\ald}$ & 1 & 0 & 0 & 0 & 0 & 0 \\
$L_{\al i}$ & 0 & 1 & 0 & 0 & 0 & 0 \\
$R_{\ald a}$ & 0 & 0 & 1 &  0 & 0 & 0 \\
$\Lambda_1^{5 - n}$ & 2 & $2 n$ & 0 & $2 - 2 n $& 0 & 0 \\
$\Lambda_2^{5 - m}$ & 2 & 0  & $2 m$  & $2 - 2 m$ & 0 & 0 \\
\hline
$X$ &  2 & 0 & 0 & 0 & 0 & 0 \\
$\scL_{ij}$ & 0 & 2 & 0 & 0 & 0 & 0 \\
$\scR_{ab}$ &0 & 0 & 2 & 0 & 0 & 0 \\
$Y_{ai}$ & 1& 1& 1&  0 & 0 & 0 \\
\hline
${1 \over \mu_1} V_{\ald i} $ & 1 & 1 & 0  & 0 &$ - 2$& 0 \\
$q^\al_\lam$ & $- 1$& 0 & 0 & 1 & 1 & 0 \\
$l^i_\lam $ & 0 & $- 1 $& 0 & 1 & 1 & 0\\
$\Lambda_2^{\prime ~ 8 - 2 n - m}$ & 4 & $2 n $& $2 m $& $4 - 2 n - 2 m$&
$- 2 n - 4$& 0 \\
\hline
$p^{\lam}_{\lamd}$ &   1 & 0 & 0 & 0 & $- 1$& 1 \\
$A^\prime_{{\lam}_1 {\lam}_2}$ &$ -2 $& 0 & 0 & 2 & 2 & $-2$\\
$v^i_{\lamd}$& $-1$& $-1 $& 0& 1& 2& 1\\
$r^a_{\lamd}$ & 0 & 0 & $-1$ & 1 & 0 & 1\\
${1\over \mu_2} G_{\lambda a}$& $-1$ & 0 & 1 & 1 & 1 & $-2$\\
$W_{ij}$&2&2&0&0&0&0\\
$\bar{\Lambda}_2^{4 n + 2 m - 10 } $ & $- 4$& $- 2 n$ & $- 2 m$ & $2 n + 2 m -
4$&
 $2 n + 4$& $4 n + 2 m - 4$\\
$\bar{\Lambda}_1^{5 - 2 m}$ & 2& 0 &$ 2 m$& $2 - 2 m$& $- 2$& $- 2 m + 2$ \\
\hline
$\mu_1$&0&0&0& 0& 2&0\\
$\mu_2$&0&0&0& 0& 0&2\\
 \hline\hline
\end{tabular}
\end{center}
\end{table}

\nc{\ib}[3]{ {\em ibid. }{\bf #1} (19#2) #3}
\nc{\np}[3]{ {\em Nucl.\ Phys. }{\bf #1} (19#2) #3}
\nc{\pl}[3]{ {\em Phys.\ Lett. }{\bf #1} (19#2) #3}
\nc{\pr}[3]{ {\em Phys.\ Rev. }{\bf #1} (19#2) #3}
\nc{\prep}[3]{ {\em Phys.\ Rep. }{\bf #1} (19#2) #3}
\nc{\prl}[3]{ {\em Phys.\ Rev.\ Lett. }{\bf #1} (19#2) #3}

\end{document}